\documentclass[a4paper]{amsart}

\usepackage[labelfont=bf, textfont=normal,
			justification=justified,
			singlelinecheck=false]{caption} 
\usepackage{graphicx} 
\usepackage{hyperref} 
\usepackage{bbm} 
\usepackage[top=3cm, bottom=3cm,
			inner=3cm, outer=3cm]{geometry}			
\usepackage{parskip} 

\title[An energy-based deep splitting method for the nonlinear filtering problem]{An energy-based deep splitting method for the nonlinear filtering problem}

\begin{document}
\author[K.~B{\aa}gmark]{Kasper B{\aa}gmark}
\address{Kasper B{\aa}gmark\\
Department of Mathematical Sciences\\
Chalmers University of Technology and University of Gothenburg\\
SE--412 96 Gothenburg\\
Sweden}
\email{bagmark@chalmers.se}

\author[A.~Andersson]{Adam Andersson} 
\address{Adam Andersson\\ 
Department of Mathematical Sciences\\
Chalmers University of Technology and University of Gothenburg\\
S--412 96 Gothenburg, Sweden\\
and Saab AB Radar Solutions\\
S--412 76 Gothenburg, Sweden} 
\email{adam.andersson@chalmers.se} 

\author[S.~Larsson]{Stig Larsson}
\address{Stig Larsson\\
Department of Mathematical Sciences\\
Chalmers University of Technology and University of Gothenburg\\
SE--412 96 Gothenburg\\
Sweden}
\email{stig@chalmers.se}

\keywords{Filtering problem, Zakai equation, stochastic partial differential equation, splitting scheme, deep learning, energy-based method}
\subjclass[2020]{60G35, 62F15, 62G07, 62M20, 65C30, 65M75, 68T07}

\maketitle
\begin{abstract}
    The purpose of this paper is to explore the use of deep learning for the solution of the nonlinear filtering problem. This is achieved by solving the Zakai equation by a deep splitting method, previously developed for approximate solution of (stochastic) partial differential equations. This is combined with an energy-based model for the approximation of functions by a deep neural network. This results in a computationally fast filter that takes observations as input and that does not require re-training when new observations are received. The method is tested on four examples, two linear in one and twenty dimensions and two nonlinear in one dimension. The method shows promising performance when benchmarked against the Kalman filter and the bootstrap particle filter.
\end{abstract}

\section{Introduction}
Nonlinear filtering is a topic intertwined between Bayesian statistics, stochastic analysis and numerical analysis. It concerns finding the conditional distribution of an unknown state, given noisy observations. There are many domains of applications for the filtering problem, e.g., finance \cite{Brigo, Date}, chemical engineering \cite{Rutzler}, weather forecasts \cite{Cassola,Duc_Kuroda}, and target tracking \cite{blackman1999design, goodman1997mathematics}. The most common approaches to approximating the nonlinear filtering problem involves extended and unscented Kalman filters as well as particle filters \cite{Kalman_Bucy,Quinn,Sarkka}. Extended and unscented Kalman filters are useful in settings with unimodal, symmetric and approximately Gaussian densities but perform poorly when dealing with multimodal models. Particle filters are useful when dealing with more complex distributions but the number of required particles, and thus the computational complexity, scales poorly in the number of dimensions of the state space \cite{Quinn}. 
The {normalized} optimal filter is known from theory to solve the Kushner--Stratonovich equation, a nonlinear Stochastic Partial Differential Equation (SPDE) \cite{Kushner}. The unnormalized optimal filter solves the Zakai equation which is a linear SPDE. These theoretically appealing facts have not been extensively used to derive practical algorithms, mostly because of the computational load associated with approximating SPDEs with classical methods such as finite element or finite difference methods. Following the recent years' extensive developments in solving PDEs and SPDEs with deep neural networks \cite{Arnulf, Arnulf_PDE, Crisan_Lobbe, Zhang_Zhou}, new opportunities arise. 
In \cite{Arnulf_PDE} the authors present a deep splitting method for high-dimensional PDEs. It relies on operator splitting and deep learning. This grid-free method was demonstrated by approximating solutions to PDEs in up to $10\,000$ dimensions. In the follow up paper \cite{Arnulf} the method was applied successfully to SPDEs in up to $50$ dimensions. In particular, it was applied to the Zakai equation for nonlinear filtering with a fixed observation sequence and a quite peculiar dynamics, chosen to admit an analytic benchmark solution.

Other deep learning based approaches that can be used to estimate the filtering density involve parameter optimization for a family of distributions, which is a common technique for estimating probability densities in regression problems \cite{Kendall_Gal, Lakshminarayanan}. 
A limitation of this approach is that it is sometimes difficult to find parameterized families of distributions that are flexible enough to fit the data well. A successful alternative to parameterizing the distributions is given by energy-based methods. This family of methods has shown excellent performance on regression problems \cite{Gustafsson_Danelljan, LeCun_Chopra, Song_Kingma}. The energy-based models are commonly trained by minimizing the negative log-likelihood. Other common techniques are based on the Kullback--Leibler divergence, noise contrastive estimation and score-matching, see \cite{Gustafsson_Danelljan, Gustafsson_Danelljan_2, Hendriks_Gustafsson, Song_Kingma}. 

In the present exploratory work we take a step towards fast and scalable nonlinear filters.
We do this by combining an energy-based approach with the deep splitting method of \cite{Arnulf} applied to the Zakai equation. In this way we use a sound probabilistic model that is flexible and does not allow negative values. A main contribution is that we allow the observation sequences as input to the model, avoiding re-training for every new sequence.
We demonstrate our model on {two} linear and Gaussian examples and on two nonlinear examples. The performance is measured with the mean absolute error, a first moment error, and the Kullback--Leibler divergence. Our method shows promising performance when benchmarked against the Kalman filter and the bootstrap particle filter. {There are no contributions to the error analysis in the present paper, however, in a future work we investigate the convergence order of this method theoretically.}

The paper is structured as follows. In Section~\ref{Background} we present a background on the filtering problem and its solution by the Zakai equation. The deep splitting method and our extension of it is derived in Section~\ref{Deep_splitt}. The energy-based approach and the full algorithm is presented in Section~\ref{Method}. It also contains a discussion of two previous approaches to solving the optimal filtering problem with deep splitting methods \cite{Arnulf, Crisan_Lobbe}. In Section~\ref{Numerics} we present our numerical results.
 
\section{Preliminaries}\label{Background}
In this section we present the notation that we use, the filtering problem, and the Zakai equation that solves it. The presentation is formal and is valid under suitable conditions. Details are omitted.

\subsection{Notation}
We denote by $\langle \cdot\,,\cdot \rangle$ and $\|\cdot\|$ the inner product and norm on the Euclidean space $\mathbb{R}^d$.
The space of functions on $[0,T]\times\mathbb{R}^{d}\to\mathbb{R}$, that are once continuously differentiable in the first variable and twice continuously differentiable in the second variable with no cross derivatives between the first variable and the second variable, is denoted $C^{1,2}([0,T]\times \mathbb{R}^d; \mathbb{R})$. Probability distributions over $\mathbb{R}^d$ are all assumed to have a probability density function with respect to Lebesgue measure. We follow the convention in Bayesian modelling to denote densities $p$ and let their arguments specify which one is meant. For a stochastic process $Y\colon \mathbb{N}\times \Omega\to\mathbb{R}^{d'}$, we denote by $Y_{k:n}$ the $d'\times (n-k+1)$-matrix $(Y_{k},Y_{k+1},\dots,Y_{n})$, where $k<n$.

\subsection{The filtering problem} \label{filtering_problem}
Filtering aims at finding the conditional distribution of an unobserved state variable, given noisy measurement of the state. In the setting of this paper both state and measurements are modeled by Stochastic Differential Equations (SDE). We denote by $(\Omega,\mathcal{A},(\mathcal{F}_t)_{0\leq t\leq T},\mathbb{P})$ a complete filtered probability space. The filtration $\mathcal{F} := (\mathcal{F}_t)_{0\leq t\leq T}$ is defined with respect to $W$ and $V$, {which are two $d$- respectively $d'$-dimensional} independent Brownian motions. For a drift coefficient $\mu\colon\mathbb{R}^d\to\mathbb{R}^d$ and a diffusion coefficient $\sigma\colon\mathbb{R}^d\to\mathbb{R}^{d\times d}$, regular enough, the process $X\colon[0,T]\times \Omega \to \mathbb{R}^d$, commonly referred to as the signal process or latent state process, is the process that for all $t\in [0,T]$, $\mathbb{P}$-a.s., satisfies
\begin{align} 
    X_t 
    &= 
    X_0 
    + \int_0^t \mu(X_s) \, \mathrm{d} s 
    + \int_0^t \sigma(X_s) \, \mathrm{d} W_s.
    \label{underlying_sde}
\end{align}
The initial condition $X_0$ is $\mathcal{F}_0$-measurable, independent of $W$ and $V$, with a distribution $p_0$.
In addition, for a sufficiently regular measurement function $h\colon\mathbb{R}^d\to\mathbb{R}^{d'}$, we define the observation process $Y\colon[0,T]\times \Omega \to \mathbb{R}^{d'}$ satisfying for all $t\in [0,T]$, $\mathbb{P}$-a.s.,
\begin{align}
    Y_t 
    &= 
    \int_0^t h(X_s) \, \mathrm{d} s 
    + V_t. 
    \label{observation_sde}
\end{align}
The filtering problem consists of finding the conditional probability density of the state $X_t$ given the observations $(Y_s)_{0\leq s \leq t}$. This is commonly referred to as the filtering density.
More precisely, the filtering density $\pi$ is at time $t$ and observation $(Y_s)_{0\leq s \leq t}$ the function, that for all measurable sets $B\subset \mathbb{R}^d$, satisfies
\begin{align*}
    \mathbb{P}(X_t\in B \mid  (Y_s)_{0\leq s \leq t})
    =
    \int_B \pi_t(x \mid (Y_s)_{0\leq s \leq t}) \, \mathrm{d} x.
\end{align*}
Since $Y$ is stochastic, in fact, $\pi_t$ is a density-valued stochastic process (under suitable regularity assumptions). {On the basis of the Kallianpur--Striebel formula, one can derive an associated SPDE whose solution is an unnormalized version of the filtering distribution. This equation is known as the Zakai equation and is first derived in weak form. Additional conditions guarantee that the distribution has a density with respect to Lebesgue measure, in which case the density, $p$, is a solution to the strong form of the Zakai equation. These derivations together with rigorous assumptions can be found in \cite{Bain_Crisan}.

Next, we present the strong form of the Zakai equation \cite{Zakai}.}
We denote an unnormalized filtering density by $(p_t)_{0 \leq t \leq T}$. To introduce the equation, we recall the differential operator $A$ associated to the process $X$ as well as its formal adjoint $A^*$, which are defined, with $a := \sigma \sigma^\top$, for $\varphi\in C^{\infty}_0(\mathbb{R}^d;\mathbb{R})$, as
\begin{align*}
    A\varphi 
    = 
    \frac{1}{2}\sum_{i,j=1}^d a_{ij}\,
    \frac{\partial^2 \varphi}{\partial x_i \partial x_j} 
    + \sum_{i=1}^d \mu_i \, 
    \frac{\partial \varphi}{\partial x_i}
    \quad 
    \text{and} 
    \quad
    A^* \varphi 
    = 
    \frac{1}{2}\sum_{i,j=1}^d 
    \frac{\partial^2}{\partial x_i \partial x_j} 
    (a_{ij}\varphi) 
    - \sum_{i=1}^d 
    \frac{\partial}{\partial x_i} 
    (\mu_i \varphi).
\end{align*}
The operator $A$ is associated to the deterministic PDE known as the Kolmogorov backward equation and the adjoint operator $A^*$ is associated to the Kolmogorov forward equation. The latter is also called the Fokker--Planck equation, which models the (unconditional) density of the signal process $X$ \cite{Oksendal}, while the Zakai equation models the (conditional) filtering density. 

The drift coefficient $\mu$, the diffusion coefficient $\sigma$, and the measurement function $h$ are assumed to be regular enough so that a solution to the Zakai equation exists for any initial density $p_0$. The strong form of the Zakai equation is to find a function $p$ such that for $t\in[0,T]$, $\mathbb{P}$-a.s.\
\begin{align}
\begin{split}
    p_t(x) 
    = 
    p_0(x) 
    + \int_0^t A^* p_s(x) \, \mathrm{d} s
    + \int_0^t p_s(x)h(x)^\top \, \mathrm{d} Y_s,
    \quad x\in \mathbb{R}^d.
\end{split}
\label{Strong_Zakai}
\end{align}
See \cite{Bain_Crisan} for details on the derivation of the equation. Following \cite{Arnulf}, we consider a more general equation of the form
\begin{align}
\begin{split} \label{SPDE_continuous}
    p_t(x) 
    =~ 
    p_0(x) 
    + \int_0^t A p_s(x) \, \mathrm{d} s 
    + \int_0^t f(x,p_s(x),\nabla p_s(x)) \, \mathrm{d} s 
    \\
    + \int_0^t b(x,p_s(x),\nabla p_s(x)) \, \mathrm{d} Y_s, 
    \quad x\in\mathbb{R}^d,
\end{split}
\end{align}
with coefficients $f\colon\mathbb{R}^d\times\mathbb{R}\times\mathbb{R}^d\to\mathbb{R}$ and $b\colon\mathbb{R}^d\times\mathbb{R}\times\mathbb{R}^d\to\mathbb{R}^{d'}$. By expanding the derivative $A^*$ we see that the Zakai equation \eqref{Strong_Zakai} is of this form with 
\begin{align*}
    f(x,u,v) 
    &= 
    \sum_{i,j=1}^d 
    \frac{\partial a_{ij}(x)}{\partial x_i }\, v_j
    + \frac{1}{2}\sum_{i,j=1}^d 
    \frac{\partial^2 a_{ij}(x)}{\partial x_i \partial x_j}\,  u 
    - \sum_{i=1}^d 
    \frac{\partial \mu_i(x)}{\partial x_i}\, u
    - 2\sum_{i=1}^d\mu_i(x) \, v_i, 
    \\
    b(x,u,v) 
    &= 
    uh(x)^\top.
\end{align*}
In addition to letting \eqref{SPDE_continuous} represent more a general SPDE, the main reason for rewriting \eqref{Strong_Zakai} in this way is that equations with leading operator $A$ admit Feynman--Kac type representations of solutions. This is used in the next section in the derivation of the deep splitting scheme. 

It is important to note that if the system of SDEs \eqref{underlying_sde}--\eqref{observation_sde} have a linear drift coefficient $\mu$, constant diffusion coefficient $\sigma$, as well as a linear measurement function $h$, then the exact density of the filter is tractable. The problem is then solved by an evolution of Gaussian densities known as the Kalman--Bucy filter, see \cite{Sarkka} for details. Otherwise there are few closed form filters, one such exception is the Beneš filter which solves a bistable model with a $\tanh$ drift. See \cite{Bain_Crisan, Benes} for details.

\section{Deep splitting method}\label{Deep_splitt}
The deep splitting method was originally derived in \cite{Arnulf_PDE} for a class of parabolic PDEs and it was extended in \cite{Arnulf} to SPDEs on the form of \eqref{SPDE_continuous}. The aim of this section is to provide a derivation of the proposed modified recursive optimization problem. In this section, we make the tacit assumption that both $f$ and $b$ are regular enough so that \eqref{SPDE_continuous} has a unique strong solution.
We start by deriving the original method, contributing with a complement to the presentation in \cite{Arnulf}, and continue by deriving an extension.
The section begins by applying a splitting scheme and continues by deriving a Feynman--Kac representation. 
Finally, we obtain a modified version of the method derived in \cite{Arnulf}, which is later used in the numerical examples. 

\subsection{A splitting scheme for the SPDE}
We introduce a partition of $[0,T]$ into $0=t_0 <t_1 <\dots < t_M = T$ and note that \eqref{SPDE_continuous} can be written for $n=0,\dots,M-1$, $\mathbb{P}$-a.s., 
\begin{align}
    \begin{split}
        p_{t}(x) = 
        p_{t_n}(x) 
        + \int_{t_n}^t A p_s(x) \, \mathrm{d} s 
        + \int_{t_n}^t f(x,p_s(x),\nabla p_s(x)) \, \mathrm{d} s \\ 
        + \int_{t_n}^t b(x,p_s(x),\nabla p_s(x)) \, \mathrm{d} Y_s, 
        \quad t\in (t_n,t_{n+1}].
    \end{split}
    \label{SPDE_discrete}
\end{align}
In the next step we split \eqref{SPDE_discrete} into two equations. {The idea behind this splitting is to treat the operator term differently than the terms involving $f$ and $b$.} We define $q$ and $\widehat{q}$ recursively on $(t_n,t_{n+1}]$, for $n=0,\dots, M-1$, as the solutions to
\begin{align}
    \begin{split} \label{nonlinear_splitt}
        {q}_{t}(x) 
        &= 
        \widehat{q}_{t_n}(x) 
        + \int_{t_n}^t f(x,{q}_s(x),\nabla {q}_s(x)) \, \mathrm{d} s 
        + \int_{t_n}^t b(x,{q}_s(x),\nabla {q}_s(x)) \, \mathrm{d} Y_s, 
        \quad t\in(t_n,t_{n+1}],
    \end{split}\\
    \begin{split}\label{linear_splitt}
        \widehat{q}_t(x) 
        &= 
        {q}_{t_{n+1}}(x) 
        + \int_{t_n}^t A \widehat{q}_s(x) \, \mathrm{d} s, 
        \quad t\in(t_n,t_{n+1}],
    \end{split}\\
    \widehat{q}_0(x) 
    &= 
    p_0(x). 
    \label{linear_splitt_2}
\end{align}
Note that $q$ and $\widehat{q}$ are piecewise smooth with respect to $t$. This splitting method is constructed such that $\widehat{q}_{t_{n+1}}\approx p_{t_{n+1}}$ for $n=0,\dots,M-1$ {and is investigated in, e.g., \cite{gyongy2003rate,gyongy2003splitting} where strong convergence order 1 in time is shown.} We further approximate \eqref{nonlinear_splitt} with an Euler--Maruyama scheme{, which in general is of strong order 0.5 in time. Merging the equations \eqref{nonlinear_splitt}--\eqref{linear_splitt}, after Euler--Maruyama approximation, into one formula we obtain the following approximation of \eqref{SPDE_discrete}, where we keep the notation of $\widehat{q}$,}
\begin{align*}
    \begin{split}
        \widehat{q}_t(x) 
        =&~ 
        \widehat{q}_{t_n}(x) 
        + \int_{t_n}^t A\widehat{q}_{s}(x) \, \mathrm{d} s 
        + f(x,\widehat{q}_{t_n}(x),\nabla \widehat{q}_{t_n}(x)) 
        (t_{n+1}-t_n) 
        \\
        &+ b(x,\widehat{q}_{t_n}(x),\nabla \widehat{q}_{t_n}(x)) 
        (Y_{t_{n+1}}-Y_{t_n}), 
        \quad t\in (t_n,t_{n+1}],
        \\
        \widehat{q}_{0}(x) =&~ p_0(x).
    \end{split}
\end{align*}
In preparation for the final algorithm we consider, for fixed $\omega\in \Omega$, deterministic input $y = (Y_{t_n}(\omega))_{n=0}^M \in \mathbb{R}^{d\times (M+1)}$ and define an approximation $\widehat p_t(x)=\widehat p_t(x,y_{0:n+1})$ recursively on $(t_n,t_{n+1}]$ for $n=0,\dots, M-1$, by
\begin{align}
    \begin{split}
        \widehat{p}_{t}(x) 
        =&~ 
        \widehat{p}_{t_n}(x) 
        + \int_{t_n}^t A\widehat{p}_{s}(x) \, \mathrm{d} s
        + f(x,\widehat{p}_{t_n}(x),\nabla \widehat{p}_{n,t_n}(x)) 
        (t_{n+1}-t_n) 
        \\
        &+ b(x,\widehat{p}_{t_n}(x),\nabla \widehat{p}_{n,t_n}(x)) 
        (y_{{n+1}}-y_{n}), 
        \quad t\in (t_n,t_{n+1}],
        \\
        \widehat{p}_{0}(x) 
        =&~ 
        p_0(x).
    \end{split}\label{splitting_V}
\end{align}
This is a splitting approximation of $p$, where the idea is to first approximate \eqref{nonlinear_splitt} with the Euler--Maruyama scheme and then solve the equation with respect to the generator $A$ exactly in the second step in \eqref{linear_splitt}. 
{There exists a unique solution $\widehat{p}$ to the Cauchy problem \eqref{splitting_V}, which belongs to $C^{1,2}((t_n,t_{n+1}]\times\mathbb{R}^d;\mathbb{R})$ \cite[Chapter 1, Theorem 10]{friedman1964partial}.}
In Subsection~\ref{Extension} this method is extended to a Milstein scheme in the special case $d=1$. In the next subsection we derive an optimization problem based on this splitting scheme.

\subsection{Derivation of a local optimization problem} \label{derivation_of_deep_splitt}
In this subsection we fix $n\in\{0,\dots,M-1\}$ and let $N\in\{n+1,\dots,M\}$ be arbitrary.  We will derive a Feynman--Kac formula for $\widehat{p}\,\vert_{(t_n,t_{n+1}]}$. 
We begin by noting that, from \eqref{splitting_V}, it is clear that 
\begin{align*}
   \begin{split}
        \frac{\partial}{\partial t}\widehat{p}_{t}(x) 
        &= 
        A\widehat{p}_t(x),
        \quad t\in(t_n,t_{n+1}].
    \end{split}
\end{align*}
Next, we reparameterize with time $t\mapsto t_N-t$, so that $t_N-t\in(t_n,t_{n+1}]$, which yields
\begin{align}
   \begin{split}
        \frac{\partial}{\partial t}\widehat{p}_{t_N-t}(x) 
        + A\widehat{p}_{t_N-t}(x) 
        = 
        0,
        \quad t\in [t_N-t_{n+1},t_N-t_n),
    \end{split}
    \label{Backward_time_derivative}
\end{align}
with final condition
\begin{align}
    \begin{split} \label{final_condition}
        \widehat{p}_{{t_N-(t_N-t_n)^-}}(x) 
        = 
        \widehat{p}_{t_n^+}(x) 
        = 
        \widehat{p}_{t_n}(x) 
        + f(x,\widehat{p}_{t_n}(x),\nabla \widehat{p}_{t_n}(x)) 
        (t_{n+1}-t_n) \\
        + \,b(x,\widehat{p}_{t_n}(x),\nabla \widehat{p}_{t_n}(x)) 
        (y_{{n+1}}-y_{n}),
    \end{split}
\end{align}
where $\widehat{p}_{t_n}$ is defined on the previous interval $(t_{n-1},t_n]$, {and we note that $t_n^+$ and $(t_N-t_n)^-$ denote a right and left limit, respectively}.
{We see that the approximation $\widehat{p}$ satisfies a Kolmogorov backward equation \eqref{Backward_time_derivative}. Such equations are studied in \cite{friedman1964partial,friedman1975stochastic}, where it is shown that there exists a unique solution $\widehat{p}\in C^{1,2}((t_n,t_{n+1}]\times\mathbb{R}^d;\mathbb{R})$, which together with its first spatial derivatives satisfies a polynomial growth bound in space. We refer to the material in Sections 4 and 5 leading up to Theorem 6.5.3 in Chapter 6 of \cite{friedman1975stochastic}.
}

We introduce a new Itô process $\widetilde{X}$ defined with respect to a $d$-dimensional Brownian motion $\widetilde{W}$ independent of $W$ and $V$. The Brownian motion $\widetilde{W}$ is adapted with respect to the filtration $\widetilde{\mathcal{F}}:= (\widetilde{\mathcal{F}}_t)_{0\leq t\leq T}$. The process $\widetilde{X}$ satisfies, $\mathbb{P}$-a.s.,
\begin{align*}
    \widetilde{X}_t 
    &= 
    \widetilde{X}_0 
    + \int_0^t \mu(\widetilde{X}_s) \, \mathrm{d} s 
    + \int_0^t \sigma(\widetilde{X}_s) \, \mathrm{d} \widetilde{W}_s, 
    \quad t\in[0,T],
\end{align*}
where $\widetilde{X}_0$ is $\widetilde{\mathcal{F}}_0$-measurable, independent of $W$, $V$ and $\widetilde{W}$, with distribution $p_0$. Since $\widetilde{X}$ is defined with respect to the generator $A$, we obtain $A$ in the integrand when applying Itô's formula. Using the fact that $\widehat{p}\in C^{1,2}((t_n,t_{n+1}]\times\mathbb{R}^d;\mathbb{R})$, Itô's formula can be applied to $\widehat{p}_{t_N-t}(\widetilde{X}_t)$ which gives $\mathbb{P}$-a.s.\
\begin{align*}
    \begin{split}
        \widehat{p}_{t_N-t}(\widetilde{X}_t) 
        = 
        \widehat{p}_{t_{n+1}}(\widetilde{X}_{t_N-t_{n+1}}) 
        &+ \int_{t_N-t_{n+1}}^t 
        \langle 
        \nabla \widehat{p}_{t_N-s}(\widetilde{X}_s),
        \sigma(\widetilde{X}_s) \, \mathrm{d} \widetilde{W}_s 
        \rangle 
        \\
        &\hspace{-4em}
        + \int_{t_N-t_{n+1}}^t 
        \Big(
        \frac{\partial}{\partial s}
        \widehat{p}_{t_N-s}(\widetilde{X}_s)
        + A\widehat{p}_{t_N-s}(\widetilde{X}_s) 
        \Big)
        \, \mathrm{d} s, 
        \quad t\in [t_N-t_{n+1},t_N-t_n).
        \\
    \end{split}
\end{align*}
Inserting \eqref{Backward_time_derivative} into the right hand side yields
\begin{align}
    \begin{split}\label{martingale_rep}
        \widehat{p}_{t_N-t}(\widetilde{X}_t) 
        = 
        \widehat{p}_{t_{n+1}}(\widetilde{X}_{t_N-t_{n+1}}) 
        +& \int_{t_N-t_{n+1}}^t 
        \langle 
        \nabla \widehat{p}_{t_N-s}(\widetilde{X}_s),
        \sigma(\widetilde{X}_s) \, \mathrm{d} \widetilde{W}_s
        \rangle.
    \end{split}
\end{align}
{The polynomial growth bounds on $\widehat{p}$, $\nabla\widehat{p}$ and the assumptions on $\sigma$ guarantee that}
\begin{align*}
    \begin{split}
        \int_{t_N-t_{n+1}}^{t_N-t_n}
        \mathbb{E}
        \Big[ \big\| 
        \sigma(\widetilde{X}_s)^{\top}\nabla
        \widehat{p}_{t_N-s}(\widetilde{X}_s)
        \big\|^2 \Big] 
        \, \mathrm{d} s 
        <\infty
    \end{split}
\end{align*}
and hence the Itô integral in \eqref{martingale_rep} is a square integrable martingale with respect to $\widetilde{\mathcal{F}}$. The conditional expectation with respect to the filtration $\widetilde{\mathcal{F}}$ gives
\begin{align}
    \begin{split}\label{conditional_zero}
        \mathbb{E}
        \bigg[
        \int_{t_N-t_{n+1}}^t 
        \langle
        \nabla \widehat{p}_{t_N-s}(\widetilde{X}_s),
        \sigma(\widetilde{X}_s) \, \mathrm{d} \widetilde{W}_s
        \rangle 
        \,\bigg|\, 
        {\widetilde{\mathcal{F}}}_{t_N-t_{n+1}}
        \bigg] 
        = 0.
    \end{split}
\end{align}
Now, as $\widetilde{X}$ is $\widetilde{\mathcal{F}}$-adapted, we can combine \eqref{martingale_rep} and \eqref{conditional_zero} to get, for $t\in [t_N-t_{n+1},t_N-t_n)$,
\begin{align}
    \begin{split}
        \mathbb{E}
        \big[
        \widehat{p}_{t_N-t}(\widetilde{X}_t)
        \,\big|\,
        \widetilde{\mathcal{F}}_{t_N-t_{n+1}}
        \big] 
        &= 
        \mathbb{E}
        \big[
        \widehat{p}_{t_{n+1}}(\widetilde{X}_{t_N-t_{n+1}})
        \,\big|\,
        \widetilde{\mathcal{F}}_{t_N-t_{n+1}}
        \big] 
        = 
        \widehat{p}_{t_{n+1}}(\widetilde{X}_{t_N-t_{n+1}}).
    \end{split} \label{reverse_conditional}
\end{align}
{We recall from the final condition \eqref{final_condition} that, for all $x\in \mathbb{R}^d$, we have,
\begin{align*}
    \widehat{p}_{t_N-t}(x) 
    \to 
    \widehat{p}_{t_n}(x),
    \quad \text{as } 
    t\uparrow (t_N-t_n),
\end{align*}
and $\mathbb{P}$-a.s.
\begin{align*}
    \widetilde{X}_t 
    \to 
    \widetilde{X}_{t_N - t_n},
    \quad \text{as } 
    t\uparrow (t_N-t_n).
\end{align*}
Combining these, we get, by the polynomial growth bounds via dominated convergence, that
\begin{align*}
    \lim_{t\uparrow (t_N-t_n)} 
    \mathbb{E}
    \Big[
    \Big |
    \widehat{p}_{t_N-t}(\widetilde{X}_t) 
    &
    -
    \widehat{p}_{t_n}  (\widetilde{X}_{t_N-t_n}) 
    + 
    f(\widetilde{X}_{t_N-t_n},
    \widehat{p}_{t_n}(\widetilde{X}_{t_N-t_n}), 
    \nabla \widehat{p}_{t_n}(\widetilde{X}_{t_N-t_n}))
    (t_{n+1}-t_n)  
    \\
    &\hspace{2em}
    +  
    b(\widetilde{X}_{t_N-t_n}, \widehat{p}_{t_n}(\widetilde{X}_{t_N-t_n}), 
    \nabla \widehat{p}_{t_n}(\widetilde{X}_{t_N-t_n}))
    (y_{{n+1}}-y_{n})
    \Big |^2
    \Big]
    = 
    0
    .
\end{align*}
Now, we take the left limit $t\uparrow (t_N-t_n)$ in \eqref{reverse_conditional} to obtain the $L^2(\Omega)$-limit}
\begin{align}
\begin{split}\label{F_conditioned_reprsentation}
    \widehat{p}_{t_{n+1}}(\widetilde{X}_{t_N-t_{n+1}}) 
    & \\
    &\hspace{-6em}
    =
    {\mathbb{E}} 
    \Big[ 
    \widehat{p}_{t_n}  (\widetilde{X}_{t_N-t_n}) 
    + f(\widetilde{X}_{t_N-t_n},\widehat{p}_{t_n}(\widetilde{X}_{t_N-t_n}), \nabla \widehat{p}_{t_n}(\widetilde{X}_{t_N-t_n})) 
    (t_{n+1}-t_n) 
    \\
    &\hspace{-4em}
    + 
    b(\widetilde{X}_{t_N-t_n}, \widehat{p}_{t_n}(\widetilde{X}_{t_N-t_n}), \nabla \widehat{p}_{t_n}(\widetilde{X}_{t_N-t_n}))
    (y_{{n+1}}-y_{n}) 
    \,\big|\, 
    {\widetilde{\mathcal{F}}}_{t_N-t_{n+1}} 
    \Big].
    \end{split}
\end{align}
To evaluate this recursion numerically we approximate the underlying stochastic $\widetilde{X}$ on the time grid $0=t_0<t_1<\dots <t_M$.
The approximation, denoted $(\widetilde{X}_n)_{n=0}^M$, with initial value $\widetilde{X}_0 \sim p_0$, is given by the Euler--Maruyama method:
\begin{align}\label{underlying_discretization}
    \begin{split}
        \widetilde{X}_{n+1} 
        &= 
        \widetilde{X}_{n} 
        + \mu(\widetilde{X}_{n})(t_{n+1}-t_n) 
        + \sigma(\widetilde{X}_{n})
        (\widetilde{W}_{t_{n+1}}-\widetilde{W}_{t_{n}}).
    \end{split}
\end{align}
For simplicity of notation, we use a uniform mesh with $t_n=n\Delta t$.  This allows us to use the same mesh for $\widetilde X$, as in the splitting method, because $t_{N-n}=t_N-t_n$ and hence $\widetilde X_{N-n}\approx \widetilde X_{t_N-t_n}$.

We exchange ${\widetilde{\mathcal{F}}}_{t_N-t_{n+1}}$ for $\mathfrak{S}(\widetilde{X}_{N-(n+1)}) \subset {\widetilde{\mathcal{F}}}_{t_N-t_{n+1}}$, with the tower property, substitute \eqref{underlying_discretization} for $\widetilde{X}$, and obtain an approximation $\overline{p}_n$ of $\widehat{p}_{t_n}$ defined by
\begin{align}
    \begin{split}
    \overline{p}_{{n+1}}(\widetilde{X}_{N-(n+1)}) 
    & \\
    &\hspace{-6em}
    =
    \mathbb{E} 
    \Big[ 
    \overline{p}_{n}  (\widetilde{X}_{N-n}) 
    + f(\widetilde{X}_{N-n},\overline{p}_{n}(\widetilde{X}_{N-n}), \nabla \overline{p}_{n}(\widetilde{X}_{N-n})) 
    (t_{n+1}-t_n) 
    \\
    &\hspace{-4em}
    +
    b(\widetilde{X}_{N-n}, \overline{p}_{n}(\widetilde{X}_{N-n}), \nabla \overline{p}_{n}(\widetilde{X}_{N-n}))
    (y_{{n+1}}-y_{n}) 
    \,\Big|\,
    \mathfrak{S}(\widetilde{X}_{N-(n+1)}) 
    \Big].
    \end{split} \label{Feynman-Kac representation}
\end{align}
{We thus introduced an additional Euler--Maruyama approximation with strong convergence order 0.5.}
This representation is an approximate Feynman--Kac formula for the solution of \eqref{Backward_time_derivative}--\eqref{final_condition} for a fixed $y = Y(\omega)$, $\omega\in\Omega$. See \cite{Oksendal} for more details on the Feynman--Kac formula and \cite{Arnulf} for more details on the derivation of this numerical scheme.

The formula in \eqref{Feynman-Kac representation} expresses the approximation $\overline{p}(\widetilde{X}_{N-(n+1)})$ as a conditional expectation with respect to $\mathfrak{S}(\widetilde{X}_{N-(n+1)})$. On the other hand, the conditional expectation can be computed using the $L^2$-minimality property \cite[Corollary 8.17]{Klenke} as a minimization problem over $L^2(\Omega,\mathfrak{S}(\widetilde{X}_{N-(n+1)}))$. With an additional argument, see \cite[Proposition 2.7]{Beck}, this can be expressed as a minimization over $C(\mathbb{R}^d;\mathbb{R})$. In the context of \eqref{Feynman-Kac representation} we have that, for $N\leq M$ and $n=0,\dots,N-1$,
\begin{align}
\begin{split}
    (\overline{p}_{{n+1}}(x))_{x\in \mathbb{R}^d} 
    =& 
    \mathop{\mathrm{arg\,min}}_{u\in C(\mathbb{R}^d;\mathbb{R})} 
    \mathbb{E} 
    \bigg[ \Big|
    u(\widetilde{X}_{N-{(n+1)}}) 
    - 
    \Big( 
    \overline{p}_{{n}}  (\widetilde{X}_{N-n})
    \\
    &\quad 
    + f(\widetilde{X}_{N-n},\overline{p}_{{n}}(\widetilde{X}_{N-n}), \nabla \overline{p}_{{n}}(\widetilde{X}_{N-n}))
    (t_{n+1}-t_n)  
    \\
    &\quad 
    + b(\widetilde{X}_{N-n},\overline{p}_{{n}}(\widetilde{X}_{N-n}), \nabla \overline{p}_{{n}}(\widetilde{X}_{N-n}))
    (y_{{n+1}}-y_{n}) 
    \Big) \Big|^2 \bigg],
    \\
    \overline{p}_0(x) =&~ p_0(x).
\end{split} \label{Minimization_recursion}
\end{align}
We recall that the solution to \eqref{Minimization_recursion} gives an approximation to \eqref{SPDE_continuous} for one fixed realization $y$.
This is the final form of the original recursive optimization problem introduced in \cite{Arnulf}. To solve and find good approximators $u\in C(\mathbb{R}^d;\mathbb{R})$ the authors of \cite{Arnulf} employ a deep learning framework to \eqref{Minimization_recursion}, hence it is called a "deep splitting method". This optimization is done in \cite{Arnulf} for each specific realization $y = Y(\omega)$. 

\subsection{Extension to non-fixed observation sequence}
{
Up to this point we have derived an optimization problem for fixed observation sequence. This requires a new training for each new observation sequence and this is very limiting in applications.
}
Instead of solving the optimization problem in \eqref{Minimization_recursion} for a fixed input sequence $y=y_{0:n+1}$, we now let the input vary over a relevant set of inputs. To achieve this, we could in principle integrate the objective in \eqref{Minimization_recursion} with respect to $y$ over some probability measure $\nu$. In practice, Monte Carlo approximation is required and for this reason it is important that $\nu$ is chosen so that relevant observation sequences are sampled. As we are interested in approximating the filtering density, the natural choice for our setting is letting $y$ be distributed according to \eqref{observation_sde}. Since $\widetilde{X}$ and $Y$ are independent, the measure $\mathbb{P} \times \nu$ can be replaced by $\mathbb{P}$ when replacing $y$ with $Y$. 
{In this way we obtain a single approximator of
the filtering density that can be used for any observation sequence, which implies that the network can be
applied to new data without re-training.}

We approximate $(X,Y)$ defined in \eqref{underlying_sde}--\eqref{observation_sde}, on the same time grid as for $\widetilde{X}$ and $\overline{p}$, with an Euler--Maruyama method. The approximations $(X_n,Y_n)_{n=0}^M$, with $X_0\sim p_0$ and $Y_0 = 0$, are defined by
\begin{align}\label{underlying_discretization_2}
    \begin{split}
        X_{n+1} 
        &= 
        X_{n} 
        + \mu(X_{n})(t_{n+1}-t_n) 
        + \sigma(X_{n})
        (W_{t_{n+1}}-W_{t_n}), 
        \\
        Y_{n+1} 
        &= 
        Y_{n} 
        + h(X_{n+1})(t_{n+1}-t_n) 
        + (V_{t_{n+1}}-V_{t_n}).
    \end{split}
\end{align}
To formalize the extended optimization problem, we want to find functions $\widetilde{p}_{{n+1}}\colon \mathbb{R}^d\times \mathbb{R}^{d\times (n+2)}\to \mathbb{R}$, for $N\leq M$ and $n=0,\dots,N-1$, satisfying
\begin{align}
\begin{split}
    (\widetilde{p}_{{n+1}}(x,y))_{(x,y)\in \mathbb{R}^d\times \mathbb{R}^{d\times (n+2)}}
    &  \\
    =
    \mathop{\mathrm{arg\,min}}_{u\in C(\mathbb{R}^d\times \mathbb{R}^{d\times (n+2)};\mathbb{R})}
    \mathbb{E}
    & \bigg[ \Big| 
    u(\widetilde{X}_{N-(n+1)},Y_{0:{n+1}})  
    -\Big( 
    \widetilde{p}_{{n}}  (\widetilde{X}_{N-n},Y_{0:{n}})
    \\
    &\hspace{-3em}
    + f(\widetilde{X}_{N-n},\widetilde{p}_{{n}}(\widetilde{X}_{N-n},Y_{0:{n}}), \nabla \widetilde{p}_{{n}}(\widetilde{X}_{N-n},Y_{0:{n}}))
    (t_{n+1}-t_n)  
    \\
    &\hspace{-3em}
    + b(\widetilde{X}_{N-n},\widetilde{p}_{{n}}(\widetilde{X}_{N-n},Y_{0:{n}}), \nabla \widetilde{p}_{{n}}(\widetilde{X}_{N-n},Y_{0:{n}}))
    (Y_{{n+1}}-Y_{n}) 
    \Big) \Big|^2 \bigg],
    \\
    &\hspace{-13.5em}\widetilde{p}_0(x,y) 
    = 
    p_0(x).
\end{split} \label{Minimization_recursion_with_obs}
\end{align}
{Solving \eqref{Minimization_recursion_with_obs} implies solving \eqref{Minimization_recursion} for almost every observation sequence $y$. This can be seen by the following argument. The objective function in \eqref{Minimization_recursion_with_obs} can be written $u\mapsto\mathbb{E}[\ell(u,\tilde X,Y)]$, where the reader can identify $\ell$. The minimum is attainable and the objective is zero at the optimum. Denoting the minimum by $u^*$ we thus have $\mathbb{E}[L(\widetilde X,Y)]:=\mathbb{E}[\ell(u^*,\widetilde X,Y)]=0$. The discrete processes $\widetilde X$ and $Y$ are independent and assuming they have densities $p_{\widetilde X}$ and $p_{Y}$ with respect to Lebesgue measure in the suitable dimensions, we have 
\begin{align*}
    \mathbb{E}[L(\widetilde X,Y)]=\int\int L(x,y)p_{\widetilde X}(x)p_{Y}(y)\,\mathrm d x \,\mathrm d y =0.
\end{align*}
This implies that
\begin{align*}
    \mathbb{E}[L(\widetilde X,y)]=\int L(x,y)p_{\widetilde X}(x)\,\mathrm d x = 0
\end{align*} 
for almost all $y$. Thus a solution to \eqref{Minimization_recursion_with_obs} solves \eqref{Minimization_recursion} for almost every $y$. }

\subsection{Accounting for a limited number of samples} \label{Extension}
\begin{figure}[b]
    \centering
    \includegraphics{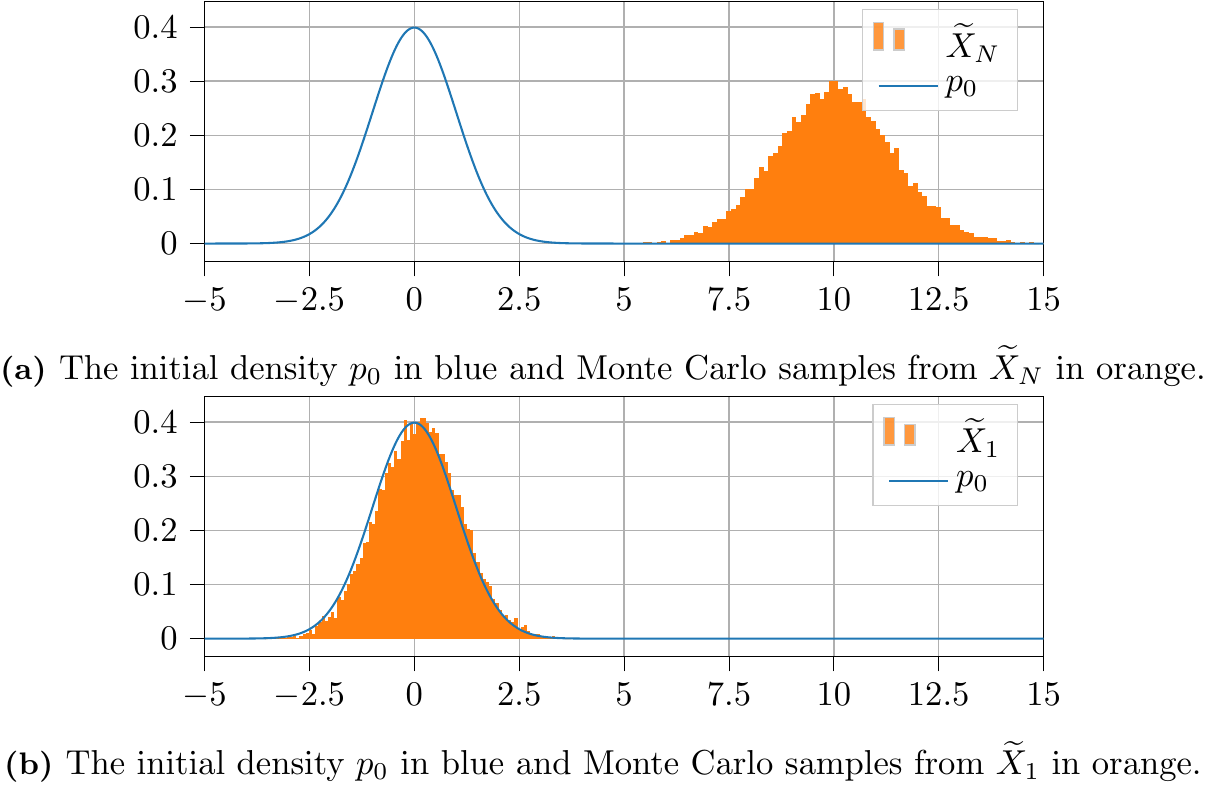}
    \caption{}
    \label{fig1}
\end{figure}

The next step is to consider Monte Carlo samples of $(Y_n,\widetilde{X}_n)_{n=0}^N$ to approximate the expectation in \eqref{Minimization_recursion_with_obs}. The limitation of having a finite number of samples can, besides classical Monte Carlo issues, in this context provide an additional problem stemming from distribution mismatch. To exemplify this, consider a one dimensional example with $p_0 = \mathcal{N}(0,1)$, $t_N = 1$, $N=100$ and a system of SDEs $(X,Y,\widetilde{X})$, in the same form as \eqref{underlying_sde}--\eqref{observation_sde}, given, for $t\in[0,1]$, $\mathbb{P}$-a.s., by
\begin{align*}
    X_t 
    &= 
    X_0 + 10t + W_t, 
    \\
    Y_t 
    &= V_t, 
    \\
    \widetilde{X}_t 
    &= 
    \widetilde{X}_0 + 10t + \widetilde{W}_t.
\end{align*}
In the first optimization step, $n=0$, the aim is to find $\widetilde{p}_{1}$ from \eqref{Minimization_recursion_with_obs}. This step consists of evaluating $p_0(\widetilde{X}_N)$ and as can be seen in Figure~\ref{fig1}a the distribution $p_0$ and the distribution of $\widetilde{X}_N$ essentially lack overlapping probability mass, being a problem in the finite sample situation of practical algorithms.
To account for this we opt for a modified approach of \eqref{Minimization_recursion_with_obs}. The setup presented so far is structured with a final time $t_N$ and having $N$ recursive minimization problems that are intertwined in the sense that the optimized models depend on training samples from time points potentially far in the future or likewise far back in time. The idea is now to use the fact that $t_N \le t_M$ is at our disposal in \eqref{Minimization_recursion_with_obs}.
What we aim to do instead is making sure each optimization problem only depends on the closest neighbours of $\widetilde{X}$ in time.
Instead of considering all time intervals of the partition simultaneously, we begin by considering $(0,t_1]$. Consider \eqref{Minimization_recursion_with_obs} with $N=1$ and $n=0$. The optimization problem is then for $\widetilde{p}_1$ defined by
\begin{align*}
\begin{split} 
    (\widetilde{p}_{{1}}(x,y))_{(x,y)\in \mathbb{R}^d\times \mathbb{R}^{d\times 2}}
    = 
    \mathop{\mathrm{arg\,min}}_{u\in C(\mathbb{R}^d\times \mathbb{R}^{d\times 2};\mathbb{R})}
    \mathbb{E} 
    & \bigg[ \Big| 
    u(\widetilde{X}_{0},Y_{0:1}) 
    - \Big(  {p}_{0}  (\widetilde{X}_{1})  
    \\
    &\hspace{-11em}
    + f(\widetilde{X}_{1},{p}_{0}(\widetilde{X}_{1}), \nabla {p}_{0}(\widetilde{X}_{1}))
    (t_{1}-t_0)  
    +  b(\widetilde{X}_{1},{p}_{0}(\widetilde{X}_{1}), \nabla {p}_{0}(\widetilde{X}_{1}))
    (Y_{{1}}-Y_0) 
    \Big) \Big|^2 \bigg] .
\end{split}
\end{align*}
By this construction we only need $X_0$ and $X_1$ from the local time interval $(0,t_1]$. In Figure~\ref{fig1}b we see how the problem of the mismatching distributions is solved, assuming our time step is sufficiently small. In this illustration the time step is $0.01$.

For the second step we consider the time interval $(t_1,t_2]$. In this step, with $n=1$, we optimize $\widetilde{p}_2$ on the samples $\widetilde{X}_{1}$ and $\widetilde{X}_{2}$ in a similar way. This is achieved by considering $N=3$ and $n=1$ in \eqref{Minimization_recursion_with_obs}.
We do this at every time interval to obtain a local optimization problem. For each time step $n$, we let $N=2n+1$ to obtain the problem on a local interval. Notice that we can choose $M$ so that we never evaluate $\widetilde{p}$ in $\widetilde{X}_m$ for $m>M$. By this construction, the obtained local minimization problem becomes for $N\leq M$ and $n=0,\dots,N-1$
\begin{align}
\begin{split}
    (\widetilde{p}_{{n+1}}(x,y))_{(x,y)\in \mathbb{R}^d\times \mathbb{R}^{d\times (n+2)}} 
    = 
    \mathop{\mathrm{arg\,min}}_{u\in C(\mathbb{R}^d\times \mathbb{R}^{d\times (n+2)})}
    \mathbb{E} 
    & \bigg[ \Big| 
    u(\widetilde{X}_{n},Y_{0:{n+1}}) 
    - \Big( \widetilde{p}_{{n}}  (\widetilde{X}_{{n+1}},Y_{0: n})  
    \\
    &\hspace{-10em}
    + f(\widetilde{X}_{{n+1}},\widetilde{p}_{{n}}(\widetilde{X}_{{n+1}},Y_{0: n}), \nabla \widetilde{p}_{{n}}(\widetilde{X}_{{n+1}},Y_{0: n}))
    (t_{n+1}-t_n)  
    \\
    &\hspace{-10em}
    + b(\widetilde{X}_{{n+1}},\widetilde{p}_{{n}}(\widetilde{X}_{{n+1}},Y_{0: n}), \nabla \widetilde{p}_{{n}}(\widetilde{X}_{{n+1}},Y_{0: n}))
    (Y_{{n+1}}-Y_{n}) 
    \Big) \Big|^2 \bigg].
\end{split} \label{Minimization_recursion_one_step}
\end{align}
By this setup we obtain the same conditions for the filtering problem and we are simply solving the optimization problems independently. This eliminates the problem previously mentioned assuming we have continuous process and a small enough time step $\Delta t$.

In \eqref{Minimization_recursion_one_step} we use the explicit Euler--Maruyama scheme. In the case $d=1$ we may instead use the Milstein scheme \cite{Kloeden_Platen}. Introducing the notation
\begin{align*}
    f_n 
    &:= 
    f(\widetilde{X}_{{n+1}},\widetilde{p}_{{n}}(\widetilde{X}_{{n+1}},Y_{0: n}), \nabla \widetilde{p}_{{n}}(\widetilde{X}_{{n+1}},Y_{0: n})), 
    \\
    b_n 
    &:= 
    b(\widetilde{X}_{{n+1}},\widetilde{p}_{{n}}(\widetilde{X}_{{n+1}},Y_{0: n}), \nabla \widetilde{p}_{{n}}(\widetilde{X}_{{n+1}},Y_{0: n})), 
    \\
    b_n^\prime 
    &:= 
    \frac{\partial}{\partial \widetilde{p}}b(\widetilde{X}_{{n+1}},\widetilde{p}_{{n}}(\widetilde{X}_{{n+1}},Y_{0: n}), \nabla \widetilde{p}_{{n}}(\widetilde{X}_{{n+1}},Y_{0: n}))
\end{align*}
the modified recursive optimization scheme reads
\begin{align}
\begin{split}
    (\widetilde{p}_{{n+1}}(x,y))_{(x,y)\in \mathbb{R}\times \mathbb{R}^{n+2}}
    = 
    \mathop{\mathrm{arg\,min}}_{u\in C(\mathbb{R}\times\mathbb{R}^{n+2};\mathbb{R})}
    \mathbb{E} & 
    \bigg[ \Big| 
    u(\widetilde{X}_{{n}},Y_{0: {n+1}}) 
    - \Big( \widetilde{p}_{{n}}  (\widetilde{X}_{{n+1}},Y_{0: n})  
    \\
    &\hspace{-14em}
    + f_n (t_{n+1} - t_n )  
    + b_n(Y_{{n+1}}-Y_{n}) 
    \hspace{-0em}
    + \frac{1}{2}b_n b_n^\prime  ((Y_{{n+1}}-Y_{n})^2- (t_{n+1}-t_n))
    \Big) \Big|^2 \bigg].
\end{split} \label{Minimization_recursion_milstein}
\end{align}

\section{The energy-based approximation scheme} \label{Method}
This section defines the proposed method for finding $\widetilde{p}_{{n+1}}$ given by the minimization problem in \eqref{Minimization_recursion_one_step} or \eqref{Minimization_recursion_milstein}. We also make a comparison to two related approaches concerned with the nonlinear filtering problem.
In particular, we consider a deep neural network $\Phi_n^{\theta_n}\colon \mathbb{R}^d\times\mathbb{R}^{d\times (n+1)}\to \mathbb{R}$, parameterized by $\theta_n$, to approximate the solution of the problem at each time step $t_n$. The optimization problem \eqref{Minimization_recursion_milstein} can be reformulated as finding $\theta^*_n$ for all $n=1,\dots,N$ satisfying
\begin{align}
\begin{split} \label{minimization_network}
    \theta^*_{1} 
    =
    \mathop{\mathrm{arg\,min}}_{\theta}
    \mathbb{E} 
    &\bigg[ \Big| 
    \Phi_1^{\theta}(\widetilde{X}_{{0}} ,Y_{0:1} ) 
    - \Big( p_0(\widetilde{X}_{{1}} )
    + f_{0} (t_{1}-t_0) 
    \\
    &\hspace{-5em}
    + b_{0}(Y_{{1}} -Y_0 ) 
    + \frac{1}{2}b_{0} b_{0}^\prime  ((Y_{1}-Y_0)^2-  (t_{1}-t_0))
    \Big) \Big|^2 \bigg], 
    \\
    \theta^*_{n} 
    = 
    \mathop{\mathrm{arg\,min}}_{\theta} 
    \mathbb{E} 
    & \bigg[ \Big|
    \Phi_n^{\theta}(\widetilde{X}_{{n-1}} ,Y_{0: {n}} ) 
    - \Big( \Phi_{n-1}^{\theta_{n-1}^*}  (\widetilde{X}_{{n}} ,Y_{0: {n-1}} ) 
    + f_{n-1} (t_{n}-t_{n-1}) 
    \\
    &\hspace{-5em}
    + b_{n-1}(Y_{{n}} -Y_{{n-1}} ) 
    + \frac{1}{2}b_{n-1} b_{n-1}^\prime  
    ((Y_{{n}} -Y_{{n-1}} )^2-  (t_{n}-t_{n-1}))
    \Big) \Big|^2 \bigg], 
    \quad n=2,\dots,N.
\end{split}
\end{align}
Analogously, we can define this optimization scheme with the Euler--Maruyama method \eqref{Minimization_recursion_one_step}, {which is used when $d\neq 1$.}
Deep neural networks have a strong ability to approximate nonlinearities and are chosen as function approximators for their ability to scale well in increasing state dimension. In this paper we construct networks with fairly simple but effective architecture. The goal of this paper and method is not necessarily to find the most effective architecture for a regression task. Instead we aim to find an effective solver and approximator of the filtering density by applying the deep learning framework to a well constructed minimization problem.

We consider an energy-based method by letting the normalized conditional density be approximated, for each pair $(x_n,y_{0:n})\in \mathbb{R}^d\times \mathbb{R}^{d\times (n+1)}$, by
\begin{align*}
    p(x_n \mid y_{0:n}) 
    \approx
    \frac{\Phi_n^{\theta_n}(x_n,y_{0:n})}{Z_n^{\theta_n}(y_{0:n})},
\end{align*}
where
\begin{align*}
    \Phi_n^{\theta_n}(x_n,y_{0:n}) 
    = 
    e^{-f_n^{\theta_n}(x_n,y_{0:n})}, 
    \quad 
    Z_n^{\theta_n}(y_{0:n}) 
    =
    \int_{\mathbb{R}^d} e^{-f_n^{\theta_n}(x,y_{0:n})}
    \, \mathrm{d} x.
\end{align*}
The main idea behind energy-based methods is to let the model output a scalar $f_n^{\theta_n}$, commonly called energy, for each input pair $(x_n,y_{0:n})$. It assigns low energy to input that is likely to occur (accurate) and high energy to unlikely input. The normalizing constant, $Z_n^{\theta_n}$, is evaluated after training to obtain the normalized density. We assume that $f_n^{\theta_n}$ is defined in a way that guarantees that $Z_n^{\theta_n}$ is finite.

\subsection{Comparison to related approaches}
Deep learning for the filtering problem is not an extensively studied topic. To the best of our knowledge, there are only two papers, \cite{Arnulf, Crisan_Lobbe}, that are based on partial differential equations. In \cite{Arnulf} a deep splitting method is used to solve the Zakai equation, while in \cite{Crisan_Lobbe} the Fokker--Planck equation is solved by deep splitting.

\subsubsection{The original approach based on the Zakai equation}
In \cite{Arnulf} the function $\overline{p}$ is approximated, for a fixed observation sequence $y_{0:N}$, by performing the minimization in \eqref{Minimization_recursion} over the parameters of a deep neural network. The model is a fully connected neural network that can take negative values which would violate the density property. 
The performance, after training, is measured by the error of the filtering density in one specific point $x\in \mathbb{R}^d$. This spatial point is selected to be close to the mode of the unnormalized density. It is shown that the model learns the value of the filter at this point with small errors at the final time step. The behavior at the other time steps is not demonstrated but in the previous work \cite{Arnulf_PDE}, in which this deep splitting method is applied to PDEs, there is a tendency towards accumulation of error. This is inherent in the recursive optimization procedure as in each step there is remaining error after optimization and in every subsequent step one optimizes with respect to a non-optimal approximation. 

In addition to allowing negative values, a problem is that the unnormalized density is only trained around typical trajectories of $\widetilde{X}$.
Elsewhere, the neural network cannot be expected to extrapolate or even have finite integral. Therefore, normalizing the approximate solution becomes meaningless and also impossible.
Finally, the networks are trained for a fixed observation sequence, making filtering in a real-time setting, often important in applications, impossible.

\subsubsection{An approach based on the Fokker--Planck equation}
In \cite{Crisan_Lobbe} the authors present a similar approach as in \cite{Arnulf}. It also involves a deep splitting scheme on an underlying equation. But in \cite{Crisan_Lobbe} this is done directly on the Fokker--Planck equation which gives the unconditional density of $X$ instead. To obtain a corresponding filter estimate, a likelihood normalization is applied after each update of the Fokker--Planck approximation. This likelihood normalization, consists of an update based on the observation $Y$ at the current time point and a normalization of the density, derived from Bayes' formula.

Similarly to the approach of the present paper and \cite{Arnulf}, the authors in \cite{Crisan_Lobbe} use a grid-free algorithm based on a Feynman--Kac type formulation, approximated by Monte Carlo simulations. The training procedure in \cite{Crisan_Lobbe} is similar to that of \cite{Arnulf} but in addition to the $L^2$-loss at each time step another term is added to encourage positive outputs of the network in the following way
\begin{align*}
    \widehat{L}(\theta) 
    = 
    L(\theta) 
    + \lambda \max{(0,-\Phi^\theta)}.
\end{align*}
In this context $L(\theta)$ represents another version of the right hand side of \eqref{minimization_network}, derived from another splitting scheme than the one in the present paper. With weight $\lambda$, negative values of $\Phi^\theta$ are penalized in order to avoid violation of non-negativity of the approximated density. This is an improvement over the use of only a scalar output as in \cite{Arnulf}, but offers no guarantees in retrieving a non-negative function. 

In \cite{Crisan_Lobbe} the authors demonstrate the method on three one-dimensional examples, two of which are solved by the Kalman filter and a third one solved by the Beneš filter. The model manages to approximate non-negative densities and the paper demonstrates corresponding errors for the means of the approximation compared to the true filters. These errors show an oscillating pattern for the more advanced examples. This means that the method is not fully consistent in its predictions of the mean. There are no numerical metrics evaluated with respect to how well the approximation captures the overall distribution, e.g., the tails and shape of the distribution. Furthermore, this work was extended in \cite{lobbe2022deep}, where the method was demonstrated on the Beneš filter with improved accuracy. The main drawback is that the model is trained for a specific observation sequence and thus would have to be trained again for new observations, similarly to the model in \cite{Arnulf}.

\section{Numerical examples} \label{Numerics}
In this section we employ the proposed method on {four} different underlying SDEs. {Two} are linear and Gaussian with the Kalman filter as benchmark. The {other two} examples are nonlinear SDEs, where we use a particle filter as benchmark. 
{Three of the examples are one-dimensional and allow us to investigate the scheme for different degrees of non-linearity for the drift, one linear and two cubic with uni-modal and bi-modal distributions, respectively. To demonstrate that the method can scale to higher dimensions we finally consider filtering of a 20 dimensional linear spring-mass system. This should be compared to \cite{Arnulf} and \cite{Arnulf_PDE} in which 50 and $10\,000$ dimensional SPDEs and PDEs were solved with the deep splitting method. We stress that our example does not exhibit symmetry (the variables of solution are not permutation invariant) as in \cite{Arnulf,Arnulf_PDE} and is therefore challenging although it has fewer dimensions. }

In Subsection~\ref{Reference_solution} we describe the reference solutions that we use as benchmarks. In Subsection~\ref{training} we describe how the model is designed and trained. Performance metrics are presented in Subsection~\ref{Metrics} and Subsection~\ref{examples} contains our numerical results. In the final part, Subsection~\ref{discussion}, we briefly discuss the model.

\subsection{Reference solution} \label{Reference_solution}
In our {four} examples we measure performance versus a benchmark solution. All our benchmarks are in discrete time, since for fixed time discretization we have a discrete system, see \cite{Lewis}. 
In the linear Gaussian case we use the Kalman Filter (KF), see, e.g., \cite{Kalman_Bucy, Sarkka}. This gives a recursive closed formula for the mean and covariance matrix of the Gaussian distribution that solves the filtering problem. Below we denote by $\mu_{\text{KF}}$ and $p_{\text{KF}}$ the mean and the density of the Kalman filter.

In the nonlinear examples we employ a Particle Filter (PF) as the benchmark. More specifically, we use a bootstrap particle filter to estimate the filtering density at each time step \cite{Sarkka}. To find a sufficient number of particles to run, we made an experiment where we looked at the variance of the estimated mean from the particle filter. This suggested using $100\,000$ particles to keep the standard deviation at approximately $0.01$. In this section we denote by $\mu_{\text{PF}}$ and $p_{\text{PF}}$ the mean and the density of the particle filter.

{Both the Kalman filter and the particle filter require knowledge of $\mu$, $\sigma$, $h$ and $p_0$ similarly to our method, making it a fair comparison.}

\subsection{Model architecture, training and evaluation} \label{training}
The regression task is one of the most common application of neural networks. In this paper we use a simple architecture consisting of a feed-forward fully connected network with 4 hidden layers with $100$ neurons each. Each layer uses batch normalization and ReLU (Rectified Linear Units) activation functions. A general overview of deep learning is found in \cite{Schmidhuber}. For details on batch normalization, see the original paper \cite{Ioffe_Szegedy} and an investigation of why this is effective in \cite{Santurkar_Tsipras}. In our energy-based approach we adapt the energy $f_n^{\theta_n}$ to each of our examples, see the corresponding Subsections~\ref{linear_example}--\ref{LSM_example}.

The optimization is performed with the ADAM optimizer together with minibatches. The latter consists of using subsets of the training data to introduce randomness of the loss function; this is a small generalization of the true Stochastic Gradient Descent method (SGD) but with more efficient iterations. Details on SGD and minibatches can be found in \cite{Goodfellow}. The ADAM optimizer makes use of the momentum of the gradient from the previous iterations; details on this can be found in  \cite{Kingma}. {We use the suggested hyperparameters $(\beta_1=0.9,\beta_2=0.999,\epsilon=10^{-8})$ from the original paper, except for the learning rate which is set to $\alpha = 10^{-5}$.}

In this paper we have access to the parameters of the underlying processes and we can generate as many samples of $(Y_n,\widetilde{X}_n)_{n=0}^N$ as we desire. Recall that $\widetilde{X}$ and $Y$ are independent. In practice it speeds up the training to limit the number of samples and reuse them in different epochs. We generate 1 million samples of each process, with the Euler--Maruyama method, and use these in minibatches to train the network. We use a rotation between different sizes of minibatches in the following order $(2^9,2^{10},2^{11},2^{12},2^{13},2^{14})$ until the loss ceases to decrease. {The rotation of sizes in minibatches creates different stochasticity in the optimizer, the larger minibatch size the less randomness. We train the network for 5-10 epochs on each batchsize before switching. The training goes on until validation error increases for 5 epochs in a row.}

After the model is trained appropriately and we have a sequence $(\widehat{\theta}_n)_{n=1}^N$, approximating $({\theta}_n^*)_{n=1}^N$, we want to estimate the normalized density and the mean of the density. In the low-dimensional case we can approximate the normalizing constant, for each observation sequence $Y_{0:n}$ and every $\widehat{\theta}_n$,
\begin{align*}
    Z_n^{\widehat{\theta}_n}(Y_{0: n}) 
    =& 
    \int_{\mathbb{R}^d} \Phi_n^{{\widehat{\theta}_n}}(x,Y_{0: n}) 
    \, \mathrm{d} x
\end{align*}
by the use of quadrature. This is not suitable in higher dimensions since these methods do not scale very well. In the high dimensional case it is more appropriate to use a Monte Carlo sampler, e.g., a Markov chain Monte Carlo such as the Metropolis--Hastings algorithm \cite{Hastings, Metropolis} {or a Hamiltonian Monte Carlo (HMC) method \cite{brooks2011handbook,duane1987hybrid}.} Given the normalization constants we can evaluate the normalized density and the corresponding mean by
\begin{align}
\begin{split}\label{approximation_method}
    p^{\widehat{\theta}_n}_n (x,Y_{0: n}) 
    &= 
    \frac{ \Phi_n^{{\widehat{\theta}_n}}(x,Y_{0: n})}
    {Z_n^{\widehat{\theta}_n}(Y_{0: n})},
    \quad x\in \mathbb{R}^d, 
    \\
    \mu^{\widehat{\theta}_n}_n(Y_{0: n}) 
    &= 
    \int_{\mathbb{R}^d} x\, p^{\widehat{\theta}_n}_n (x,Y_{0: n}) 
    \, \mathrm{d} x.
\end{split}
\end{align}
{In the high-dimensional case we make use of an HMC sampler to find the mean and normalizing constant of the approximation. Specifically, we use a step size of $0.1$ and a final time of $1.0$ in the leapfrog step of the sampler. More details on this method can be found in \cite{brooks2011handbook}.}
\subsection{Metrics} \label{Metrics}
In the previous section, finding the (unnormalized) density was emphasized. However, the most common estimates in filtering problems are the mode and the mean. In cases where the filtering density is symmetric and unimodal, or even Gaussian, these two coincide. In other applications one can argue about which is most relevant but in this context we opt to primarily measure the mean of the distribution.

In order to approximate the expectation of different metrics that we want to evaluate, we simulate $M$ coupled state-observation sequences $(X,Y)^{(m)} = (X_{n}^{(m)},Y_{n}^{(m)})_{n=0}^N$ for $m=1,\dots,M$.
We calculate the mean $\mu_n^{(m)}$ and the density $p_n^{(m)}$ of the reference solution for each time step $t_n$, and sample index $m$. Similarly, we denote the normalized density and mean of a generic approximation by
\begin{align*}
    \begin{split}
        \widehat{p}^{(m)}_n(x) 
        &= 
        \widehat{p}_n (x,Y_{0: n}^{(m)}),
        \quad x\in \mathbb{R}^d, 
        \\
        \widehat{\mu}^{(m)}_n 
        &= 
        \widehat{\mu}_n (Y_{0: n}^{(m)}).
    \end{split}
\end{align*}
In the examples we evaluate these with our method and also with two standard methods that we consider as baselines for comparison. The first way of using the mean is to directly measure the {Euclidean} distance between the empirical mean of the approximation and the mean from the reference solution. We call this error the First Moment Error (FME). This {metric is defined by}
\begin{align}\label{mean_error}
    \text{FME} 
    = 
    \frac{1}{M}\sum_{m=1}^M \big\| \mu_n^{(m)} - \widehat{\mu}_n^{(m)} \big\|,
    \quad \text{for } n=1,\dots,N.
\end{align}
The second way is to compare the mean directly with the true state $X_{n}$ at each time step. This is a common measurement of how well an approximation performs \cite{Bai, Xu, Hendriks_Gustafsson, Masouri}. Let $\mathfrak{m}_n^{(m)}$ denote the mean, either from the true filter or from an approximation. This measurement is known in the literature as the Mean Absolute Error (MAE). However, it is not an actual error since we do not expect it to converge to zero. More precisely, in the evaluation we aim to achieve almost the same MAE value from our approximation as from the true filter. The metric is defined by
\begin{align}\label{state_vs_true}
    \text{MAE} 
    = 
    \frac{1}{M}\sum_{m=1}^M \big\|X_{n}^{(m)} - \mathfrak{m}_n^{(m)}\big\|, 
    \quad \text{for } n=1,\dots,N.
\end{align}
The final, and perhaps the most interesting measurement, is the Kullback--Leibler divergence between the entire filtering density and the one generated with our method. This shows how well we manage to capture the density in the whole domain. The distance consists of taking the expectation of the difference between the logarithms of the two distributions, with respect to one of the distributions. It is important to note that the divergence is not a metric since it is not symmetric. One can consider both the forward divergence, in which one takes the expectation with respect to the true distribution $p$, and the reverse divergence by taking the expectation with respect to the approximation $\widehat{p}$. The forward divergence is considered mean seeking and the reverse divergence is considered mode seeking \cite{Zhang}. In this paper we consider the forward divergence averaged over $M$ samples. We sample $x_{n}^{(k,m)}$ from $p_n^{(m)}$ for $k=1,\dots,K$ and evaluate the averaged Kullback--Leibler Divergence (KLD) $D_{\text{KL}}$ according to
\begin{align}\label{KL_div}
    \text{KLD} 
    = 
    D_{\text{KL}}(p_n\|\widehat{p}_n) 
    &= 
    \sum_{m=1}^M\sum_{k=1}^K
    \log\Bigg(
    \frac{p_n^{(m)}
    \big(x_{n}^{(k,m)}\big)}
    {\widehat{p}_n^{(m)}
    \big(x_{n}^{(k,m)}\big)}
    \Bigg),
    \quad \text{for } n=1,\dots,N.
\end{align}
In \cite{Yeo} this distance is used to measure how well the distributions match. It is also common to use the Kullback--Leibler divergence to calculate likelihood ratios in particle filters \cite{Masouri}. It can also be used directly as a loss function during training, such as in \cite{Gustafsson_Danelljan_2}, where it is used for unconditional densities in a data driven manner. 

In the examples in the next subsection we evaluate these metrics between our proposed approximation, which we denote by the Energy-Based Deep Splitting (EBDS), and the reference solution. For comparison we also evaluate these metrics on other approximate solutions. For the nonlinear examples we employ an Extended Kalman Filter (EKF) as a baseline \cite{Sarkka}. We expect the EKF to yield decent estimates when the filtering density is unimodal. In the linear examples we use particle filters with fewer particles as baseline. These are less exact but are the most commonly used tool and thus an interesting comparison to our model.

\subsection{Examples}
\label{examples}
In this subsection we test the performance of our model on three different underlying SDEs with $d=1$ {in Sections \ref{linear_example}--\ref{bimodal_example}, and one with $d=20$ in Section \ref{LSM_example}. For the linear examples we} benchmark the model against the Kalman filter which provides the true solution. In the nonlinear examples we use the bootstrap particle filter. To simplify the comparison between the {four} examples we use the same constant time step $\Delta t = 0.01$ in all examples. In the first two we have $N=100$ and final time $t_N = 1$, while in the last two, we have $N=50$ and $t_N=0.5$. In all {one-dimensional} examples the measurement function in \eqref{observation_sde} is linear and defined by
\begin{align*}
    h(x) 
    = 
    \beta x,
    \quad x\in \mathbb{R}.
\end{align*}
We set $\beta = 1$ and it is worth noting that this results in a very large observation noise. In \cite{Crisan_Lobbe} they have two linear examples with a factor $\beta =90$ in the measurement function $h$, resulting in much smaller observation noise. For a similar setting, in \cite{Xu} they opt for a factor $\beta = 5.5$ in most of their examples. Furthermore, we consider the same diffusion coefficient for $X$ in all {one-dimensional} examples, given by
\begin{align*}
    \sigma(x) 
    = 
    1,
    \quad x\in \mathbb{R}.
\end{align*}
Finally, we consider an initial density $p_0 = \mathcal{N}(0,1)$ in all {one-dimensional} examples.
\subsubsection{Mean-reverting linear state equation} \label{linear_example}
Here we consider an underlying process \eqref{underlying_sde} with drift coefficient defined by
\begin{align*}
    \mu(x) 
    = 
    -x, 
    \quad x\in \mathbb{R}.
\end{align*}
The solution to \eqref{underlying_sde} is an Ornstein--Uhlenbeck process. This process is mean reverting towards $0$ and this can be seen in the underlying density. The density at time $t$, conditioned on $X_0 = x_0$, is given by  $\mathcal{N}(x_0e^{-t},\frac{1}{2}(1-e^{-2 t}))$. With the prior $x_0\sim \mathcal{N}(0,1)$, the posterior density is given by $\mathcal{N}(0,\frac{1}{2}(1+e^{-2t}))$. 

{In this linear example we use a neural network $\widetilde{f}^{\widehat{\theta}_n}_n \colon (x_n,y_{0:n}) \mapsto (x_n,\xi_1,\xi_2) \in \mathbb{R}^3$ and concatenate it with a layer specifically designed for the problem, 
\begin{align} \label{linear_architecture}
    g(x_n,\xi_1, \xi_2) = \xi_1 + (x_n-\xi_2)^2 \mathbbm{1}_{\mid x_n\mid >\xi_2}.
\end{align}
The energy function is then defined as $f^{\widehat{\theta}_n}_n = g \circ \widetilde{f}^{\widehat{\theta}_n}_n$.} The second term is added to guarantee that the density is integrable with essentially Gaussian tails. {In this way we build structure into the model, based on prior knowledge of the problem. This is further discussed in Section \ref{discussion}.}

We employ our method and present the different metrics in the first column of Figure~\ref{all_results} over the 100 time steps. In Figure~\ref{all_results}a and Figure~\ref{all_results}d, we see the (unconditional) density of $X$ and an illustration of a trajectory as well as the corresponding filter mean and estimated mean. In this example we also demonstrate the performance of a particle filter with $1000$ (PF-1000) particles as a baseline for comparison. In Figure~\ref{all_results}g we present the MAE \eqref{state_vs_true} with $\mathfrak{m}$ given by the Kalman filter as well as with our method and PF-1000. In Figure~\ref{all_results}j we see the FME, the difference between the true mean of the Kalman filter and the empirical mean of our model, measured as in \eqref{mean_error} with $\mu = \mu_{\text{KF}}$. We also present the FME between the mean from PF-1000 and the true mean. Finally in Figure~\ref{all_results}m the KLD \eqref{KL_div} is presented. We can see that the model performs well with respect to the MAE but slowly lose accuracy over time. Similarly the FME and KLD show increasing error over time but decent results compared to PF-1000. In the example trajectory in Figure~\ref{all_results}d we can see that the mean of our method follows the true mean very closely in the beginning but loses accuracy toward the final time.

We present our approximation for one arbitrarily chosen observation sequence in Figure~\ref{linear_densities}. This illustration was inspired by the figures in \cite{Crisan_Lobbe}. In Figure~\ref{linear_densities}a we see the time evolution of the density in blue. On the right, in Figure~\ref{linear_densities}b--\ref{linear_densities}d, we see snapshots at three different times of the density compared to the true density given by the Kalman filter.

\subsubsection{Mean-reverting nonlinear state equation}\label{mr_example}
Here we consider an underlying SDE \eqref{underlying_sde}, where the drift coefficient is defined by
\begin{align*}\label{mildnonlinear_example}
    \mu(x) 
    = 
    -x - x^3,
    \quad x\in\mathbb{R}.
\end{align*}
This process has roughly a similar statistics as the linear process of Subsection~\ref{linear_example}. More precisely, we have a process that is mean reverting towards the long term mean $0$. The cubic term increases the strength of the mean reversion, compared to the Ornstein--Uhlenbeck process. This results in a more narrow underlying distribution with smaller tails. One could argue that this problem constitutes a setting for which the EKF is suitable since the distribution is unimodal and symmetric, and hence we use it as a baseline comparison. We use a neural network with the same structure as for the previous example, defined in \eqref{linear_architecture}.

We demonstrate our method in the second column of Figure~\ref{all_results}. The figures follows the same format as for the linear example. In Figure~\ref{all_results}b wee see the density of $X$ at the final time $T=1$. In Figure~\ref{all_results}e we see one trajectory of $X$ together with the corresponding filter estimates. For the metrics MAE, FME and KLD we use a bootstrap particle filter with 100\,000 particles as our benchmark. In addition to evaluating the metrics for our method, we also demonstrate the performance of the EKF. These errors are presented in Figure~\ref{all_results}h, \ref{all_results}k and \ref{all_results}n. Clearly, in all three metrics our model performs better than the EKF. It is also interesting to note that in this example we reach a steady error level over time in both FME and KLD.

Similarly to the previous example we present the approximated density in the whole domain for a single observation sequence in Figure~\ref{mildnonlinear_densities}. In these figures the reference solution is given by a particle filter, presented in the snapshots of Figure~\ref{mildnonlinear_densities}b--\ref{mildnonlinear_densities}d, in red.

\subsubsection{Bistable nonlinear state equation} \label{bimodal_example}
In our final example we consider an underlying SDE \eqref{underlying_sde} with the drift coefficient
\begin{align*}
    \mu(x) 
    = 
    \frac{2}{5}(5x - x^3),
    \quad x\in\mathbb{R}.
\end{align*}
Compared to the mean reverting setting, the two terms have different signs here. This results in two attracting equilibria, positioned symmetrically around $0$, creating a bimodal underlying distribution. Compared to the mean reverting example we expect the extended Kalman filter to perform poorly in this setting. 

{In this bimodal example we instead let the neural network be defined by $\widetilde{f}^{\widehat{\theta}_n}_n \colon (x_n,y_{0:n}) \mapsto (x_n,\xi_1,\xi_2,\xi_3,\xi_4,\xi_5) \in \mathbb{R}^6$ and concatenate it with a different layer $g$ defined as
\begin{align*} 
    g(x_n,\xi_{1:5}) =
    \xi_1 
    + \xi_2(x_n-\xi_3)^2 \mathbbm{1}_{x_n<\xi_3} 
    + \xi_4(x_n-\xi_5)^2 \mathbbm{1}_{x_n>\xi_5}.
\end{align*}
The energy function is defined as before by the concatenation $f^{\widehat{\theta}_n}_n = g \circ \widetilde{f}^{\widehat{\theta}_n}_n$.} The second and third term guarantee that the density is integrable. Compared to the architecture for the linear example \eqref{linear_architecture}, we let the exponential decay from these extra terms be non-symmetrical around $0$. We have also introduced $\xi_2$ and $\xi_4$ to increase the flexibility of the model; both of these are defined as non-negative outputs to guarantee that the model does not explode.

\begin{figure}
    \centering
    \includegraphics[scale=0.95]{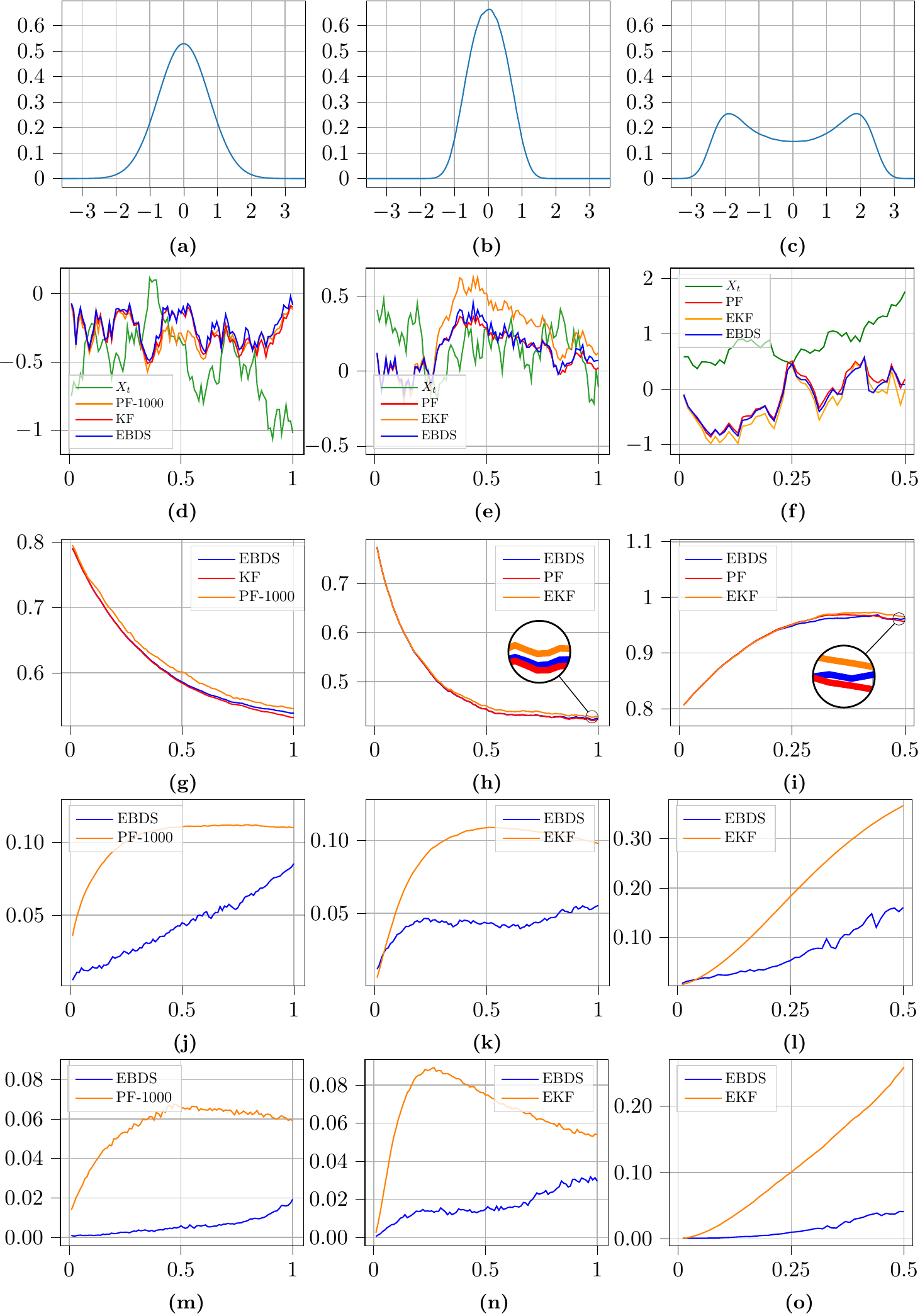}
    \caption{The figure presents numerical results for the three examples with respect to time. Left to right: Linear, mean reverting and bistable example.
    Top to bottom: Underlying densities of $X_T$, example trajectories, MAE, FME and KLD. Our method (EBDS) is illustrated in blue, the reference solution in red and a baseline in orange.}
\label{all_results}
\end{figure}

\begin{figure}
    \centering
    \includegraphics{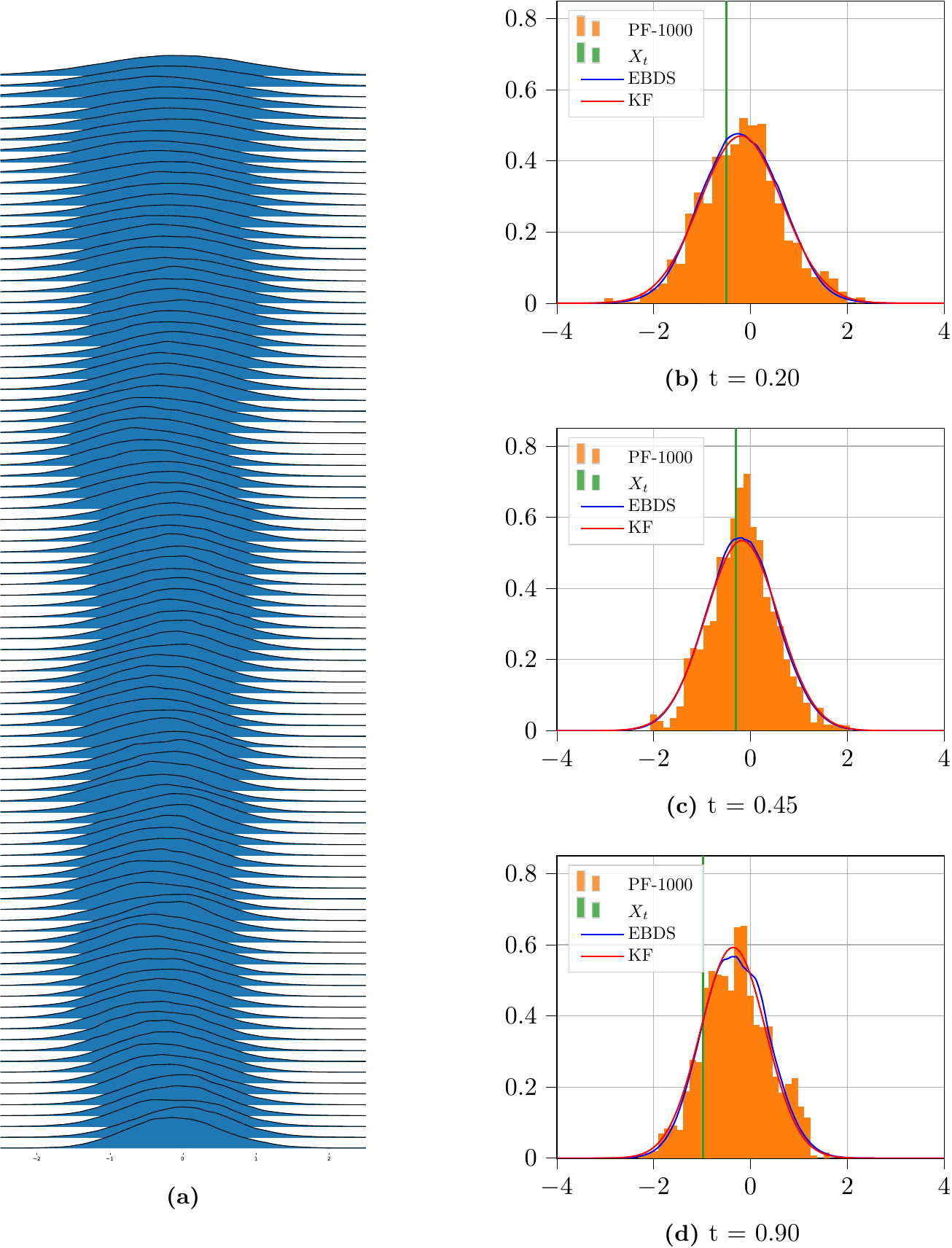}
    \caption{In (a) we see the time evolution of the density given by our model in blue, {from time $t=0.01$ at the top to $t=1.00$ at the bottom}. (b)--(d) are snapshots of the true filtering density given by the KF in red, the density from our model (EBDS) in blue, the PF-1000 in orange as well as the true state $X_t$ in green.}
    \label{linear_densities}
\end{figure}

\begin{figure}
    \centering
    \includegraphics{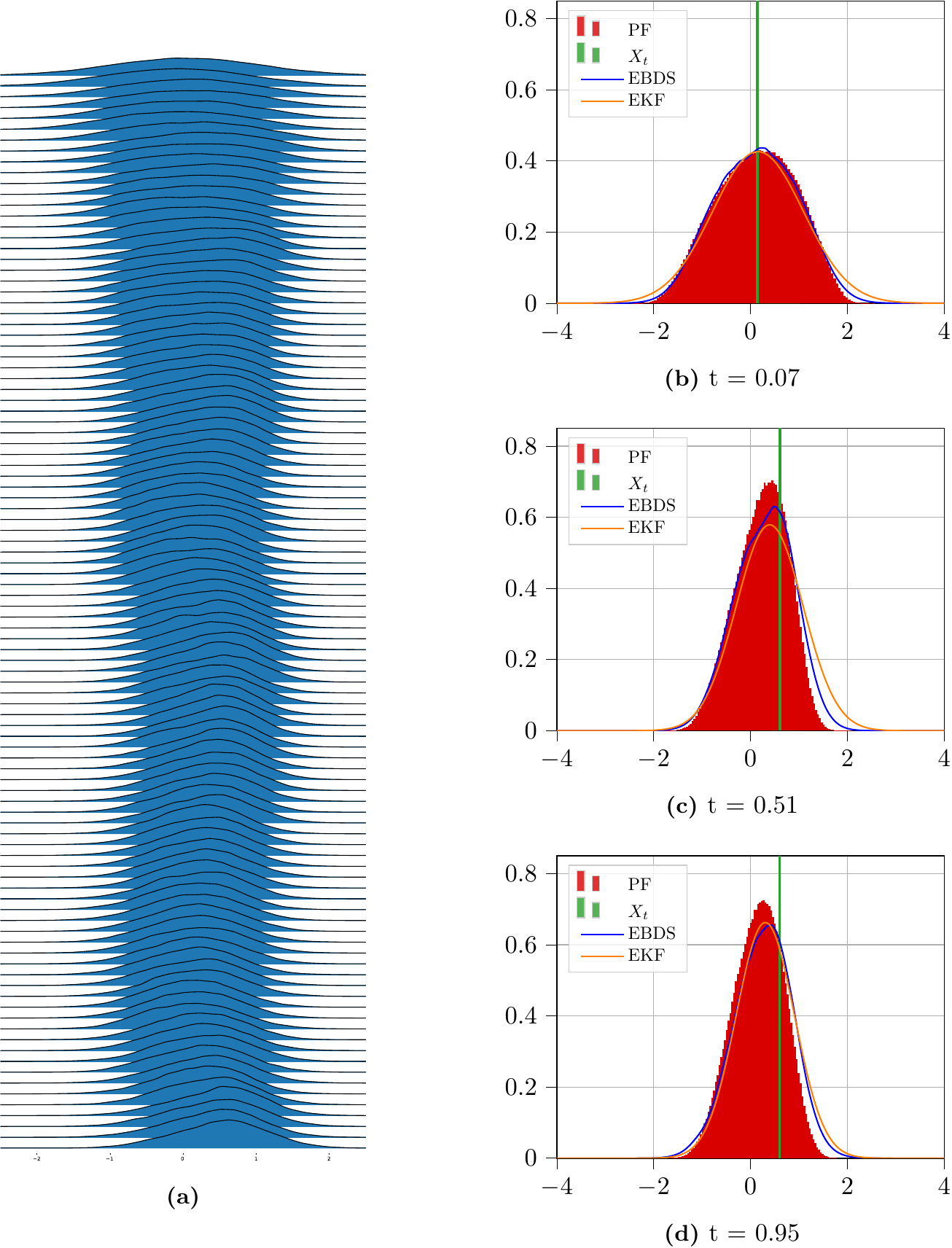}
    \caption{In (a) we see the time evolution of the density given by our model in blue, {from time $t=0.01$ at the top to $t=1.00$ at the bottom.}
    (b)--(d) are snapshots of the true filtering density given the PF in red, the density from our model (EBDS) in blue and the EKF in orange as well as the true state $X_t$ in green.}
    \label{mildnonlinear_densities}
\end{figure}

In the right column of Figure~\ref{all_results} we present the metrics of our method employed on this bistable SDE. We compare the performance of our model to that of the EKF which we believe to yield poorer approximations in this setting when the filtering density might not be unimodal. In Figure~\ref{all_results}c, we see the density of $X$ at the final time $T=0.5$. In Figure~\ref{all_results}i, \ref{all_results}l and \ref{all_results}o we see that our method outperforms the EKF in all metrics. In the MAE we see a similar performance for our model as for the PF. We also note that the FME and KLD increase almost linearly with time.

\begin{figure}
    \centering
    \includegraphics{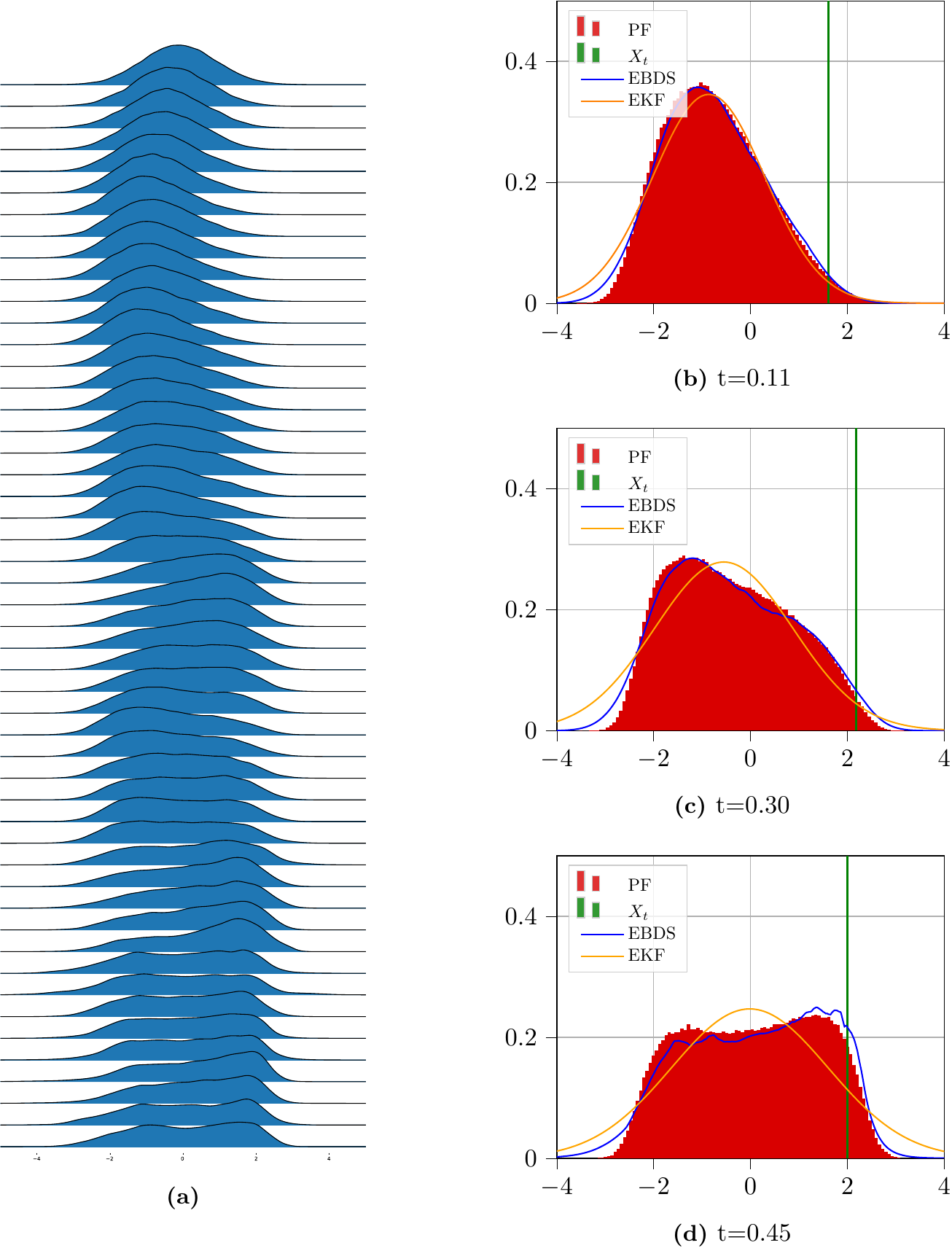}
    \caption{In (a) we see the time evolution of the density given by our model in blue, {from time $t=0.01$ at the top to $t=0.50$ at the bottom.} (b)--(d) are snapshots of the true filtering density given by the PF in red, the density from our model (EBDS) in blue and the EKF in orange as well as the true state $X_t$ in green.}
    \label{bimodal_densities}
\end{figure}

Similarly to earlier examples we illustrate the approximation of the filtering density over time for a single observation sequence. This is seen in Figure~\ref{bimodal_densities}, where on the right we see snapshots of the approximation as well as the reference given by the PF. In the last snapshot we can see how our method starts to lose accuracy when compared to the PF but still achieves favorable result to the EKF. 

\begin{figure}[ht]
\captionsetup{justification=raggedleft}
\includegraphics{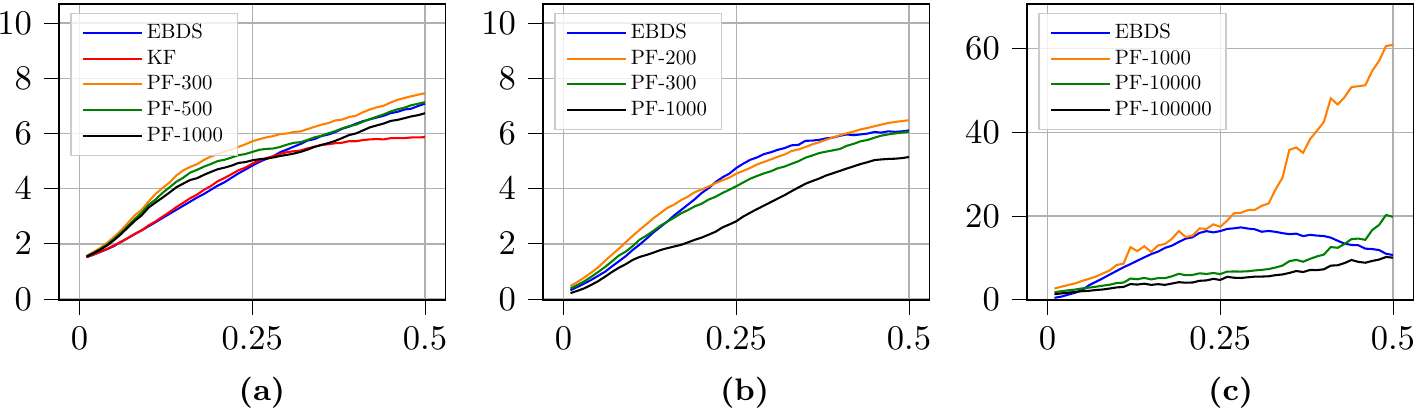}
\vspace{-10pt}
\captionsetup{justification = justified}
\caption{The figure presents numerical results for the LSM example. Left to right: MAE, FME and KLD. Our method (EBDS) is shown in blue, the reference solution (KF) in red, and particle filters with different accuracies in orange, green and black.}
\label{LSM_figure}
\end{figure}

{
\subsubsection{Linear spring-mass}
\label{LSM_example}
In this example we consider filtering of a high-dimensional equation.
Consider a mechanical system with $M$ masses connected in series with springs and dampers. They move friction-less and without gravity. The state of the system consists of displacements (from equilibrium points) and velocities. The spring forces are linear and the resulting system of ordinary differential equations is linear. We denote the masses $m_i$, $i=1,\dots,M$, the stiffness and damping constants, $k_i$ and $c_i$, $i=1,\dots,M+1$, respectively. In this example we consider a system perturbed by noise, e.g., stemming from vibrations. This results in a $2\times M$ dimensional SDE where the first $M$ dimensions represent the displacements and the last $M$ dimensions represent the velocities. The equation is given by
\begin{align*} 
    X_t 
    &= 
    X_0 
    + \int_0^t 
    \begin{bmatrix}
        0_{M\times M} & I_{M\times M} \\
        A_{21} & A_{22}
    \end{bmatrix}
    X_s \,\mathrm{d} s 
    + \int_0^t 
    \begin{bmatrix}
        \sigma_{1}I_{M\times M} & 0_{M\times M} \\
        0_{M\times M} & \sigma_2 I_{M\times M}
    \end{bmatrix} 
    \,\mathrm{d} W_s,
\end{align*}
where
\begin{align*}
    A_{21} 
    &= 
    \begin{pmatrix}
    -\frac{k_1+k_2}{m_1} & \frac{k_2}{m_1} & & \\
    \frac{k_2}{m_2} & \ddots & \ddots & \\
    & \ddots & \ddots & \frac{k_M}{m_{M-1}}\\
    &  & \frac{k_M}{m_M} & -\frac{k_M + k_{M+1}}{m_M}\\
    \end{pmatrix}
    ,\quad
    A_{22} 
    &= 
    \begin{pmatrix}
    -\frac{c_1+c_2}{m_1} &  & \\
     & \ddots & \\
    &  & -\frac{c_M+c_{M+1}}{m_{M}}
    \end{pmatrix}.
\end{align*}
In our example we consider identical masses, $m_i = 1$, stiffness constants $k_i = 5$ and damping constants $c_i=0.1$. We also choose the diffusion constants $\sigma_1 = \sigma_2 = 1$. Thus we have the same level of noise here as in the previous examples. Note that in the setting of this paper we have $d=2M$ and we consider a selection of linear measurements of displacements and velocities. 
We let the number of masses be $M=10$, and we define an observation process $Y$ in \eqref{observation_sde}, with $d'=10$, by a measurement function $h:\mathbb{R}^{20}\to \mathbb{R}^{10}$ with $h(x) = Hx$, where $H$ is the $10\times 20$ matrix with zero entries except for $H_{i,2i-1} = H_{j,2j} = 1$ for $i=1,2,3,4,5$ and $j=6,7,8,9,10$. In other words, we measure five positions and five velocities. 

For this example we used a different architecture for the model. Let the input $x_n$ to the network be split as $x_n = (x_n^{\text{pos}}, x_n^{\text{vel}})$, where the two parts represent the 10-dimensional vectors of displacements and velocities, respectivly.
Let 
$\widetilde{f}^{\widehat{\theta}_n}_n \colon(x_n,y_{0:n}) \mapsto (x_n^{\text{pos}},x_n^{\text{vel}},\alpha,\xi_1,\xi_2,\beta_1,\beta_2,\lambda_1,\lambda_2)$ define the model, where $(\alpha,\xi_1,\xi_2,\beta_1,\beta_2,\lambda_1,\lambda_2)\in \mathbb{R}^{1\times M\times M\times 1 \times 1\times 1\times 1}$. 
The intuition behind this model is that $\alpha$ fits the training data with high flexibility. The other outputs are defined to handle domains outside of the support of the training data similarly to the previous examples. More precisely $(\xi_1,\xi_2)$ define means and $(\beta_1,\beta_2)$ variances for the Gaussian tails outside some domain defined by $\lambda_1$ and $\lambda_2$. The network $\widetilde{f}^{\widehat{\theta}_n}_n$ is concatenated with 
\begin{align*} 
    g(x_n^{\text{pos}},x_n^{\text{vel}},&
    \alpha,\xi_1,\xi_2,\beta_1,\beta_2,\lambda_1,\lambda_2) 
    =
    \\
    &\alpha 
    + \beta_1 \|x_n^{\text{pos}} - \xi_1 \|^2
    \mathbbm{1}_{\|x_n^{\text{pos}} - \xi_1 \|^2>\lambda_1} 
    + \beta_2\|x_n^{\text{vel}} - \xi_2 \|^2
    \mathbbm{1}_{\|x_n^{\text{vel}} - \xi_2 \|^2>\lambda_2}.
\end{align*}
The energy function is then defined as $f^{\widehat{\theta}_n}_n = g \circ \widetilde{f}^{\widehat{\theta}_n}_n$. In this example we construct the network based on knowledge of the problem and let the model learn different tail properties for the densities of the positions and of the velocities.

In Figure \ref{LSM_figure} we present the metrics of our method employed on this linear spring-mass example. We compare it directly to the Kalman filter which provides the true solution. As in the previous examples we demonstrate the performance with particle filters with fewer particles. This time we employ three different accuracies to understand the performance by comparisons. By doing this we encapsulate the performance of our method for each metric. The performances in MAE and FME are, considering the low number of particles in the comparisons, less satisfactory than for the one-dimensional examples. The performance KLD on the other hand is much better, in particular for the final time.  
}

\begin{figure}[ht]
\captionsetup{justification=raggedleft}
\includegraphics{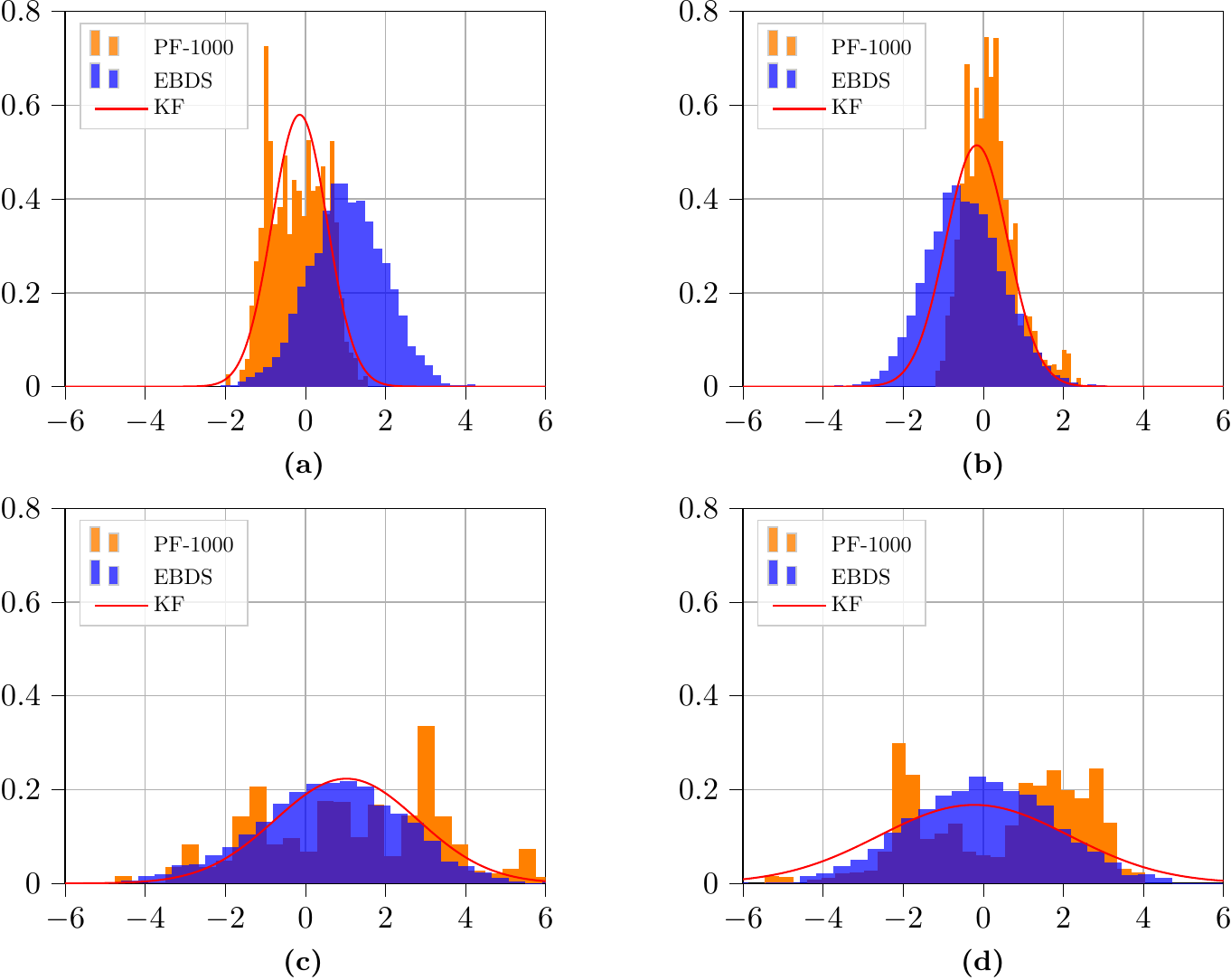}
\vspace{-5pt}
\captionsetup{justification = justified}
\caption{The figure shows the marginal densities, for a single observation sequence, of the Kalman filter in red, our method (EBDS) in blue and a particle filter with $1000$ particles in orange. The left column shows dimensions that have been observed and the right column shows dimensions which are not observed. The top row contains positions and the bottom row shows velocities.}
\label{LSM_density}
\end{figure}

{To also evaluate the method qualitatively, we compare the marginal densities for a single observation sequence for the different methods.
In Figure \ref{LSM_density} we illustrate four different marginal densities of our method compared to the KF and a PF with $1000$ particles. This is to demonstrate some of the qualities of the model and should not be used as a reference of how well the method performs more generally. Clearly the model manages to learn Gaussian tails outside of the domain of the training data. The model is also flexible enough to match the variance of the Kalman filter.
In this figure one can also observe that the model fits the velocities slightly better than the positions. This behaviour was observed for most observation trajectories, which is not shown here.
}

\subsection{Discussion} \label{discussion}
The approximation method depends on the splitting of the SPDE \eqref{SPDE_discrete}. This is a crucial step in the derivation and we have not analysed the error caused by this approximation. Furthermore, the approximation of $q$ in \eqref{nonlinear_splitt} and the stochastic processes $(X,Y,\widetilde{X})$ yield discretization errors. 
{This means that} in each local optimization step we commit an error, even if a global optimum was to be found.
Looking at the Kullback--Leibler divergence for the three one dimensional examples, we see roughly a linear increase in the error over time. Similar behaviour was observed in \cite{Arnulf_PDE} for the approximation of PDEs. 
We believe that this is mostly due to the accumulation of local errors from each time step $t_n$. 
{In a second part of this study we will do an error analysis of the different approximations employed here.}

For the first few time steps the training was a simple task and convergence was fast. This is likely due to the simplicity of the problem, i.e., the filtering distribution has not yet deviated much from the given initial distribution, and also the observations are fewer. At later time steps, especially for the bistable example, the sought density looks very different for each observation sequence. For some sequences the density is unimodal throughout every time step, while for others it might create a bimodal density such as the one seen toward the final time in Figure~\ref{bimodal_densities}a. In the bistable example we chose to stop at $N = 50$ time steps because of the increased difficulty of training the models. {Similar difficulty was found in the linear spring-mass example.} In particular, the networks were repeatedly retrained until a satisfactory solution was found. 

{In the different examples we also incorporated additional architecture into the model adapted to the dynamics of the particular example. The main idea with this is to guarantee that we obtain a function that can be normalized by letting the function go to 0 outside of the support of the training data.
Researchers who seek a problem-agnostic method might find this unsatisfactory, but we find it a sound approach to understand the problem at hand and utilize structure from it, known from theory or gained from simulations. 
}

{The purpose of our development of the method is to get a filter that scales better than particle filters. While the latter are performing a better inference with sufficiently many particles, the neural network of EBDS, after training, are orders of magnitudes faster and there is a marginal difference in computational time for 1 or 20 dimensions. This was also demonstrated in \cite{Arnulf,Arnulf_PDE}. In fact, the number of particles required for a particle filter scales exponentially in the state dimension \cite{snyder2015performance}. Our work is a first small step towards non-linear filters in high dimensions, based on partial differential equations and applicable in a real time setting.  
}

\subsection*{Acknowledgements}
We are grateful to Oskar Eklund, Moritz Schauer and Kristoffer Andersson for useful input during the writing of this paper.
The work of K.B. and S.L. was supported by the Wallenberg AI, Autonomous Systems and Software Program (WASP) funded by the Knut and Alice Wallenberg Foundation. 

\bibliographystyle{abbrv}
\bibliography{bib}

\end{document}